\pdfoutput=1
\documentclass[twocolumn,pra,superscriptaddress,longbibliography]{revtex4-1}
\usepackage{graphicx,amsmath,amssymb}
\usepackage[colorlinks=true, citecolor=blue, urlcolor=blue, linkcolor=red]{hyperref}

\begin{document}
\title{Quantum speed limit of a noisy continuous-variable system}
\author{Wei Wu}
\affiliation{Lanzhou Center for Theoretical Physics, Key Laboratory of Theoretical Physics of Gansu Province, Lanzhou University, Lanzhou 730000, China}
\author{Jun-Hong An}
\email{anjhong@lzu.edu.cn}
\affiliation{Lanzhou Center for Theoretical Physics, Key Laboratory of Theoretical Physics of Gansu Province, Lanzhou University, Lanzhou 730000, China}

\begin{abstract}
Setting the minimal-time bound for a quantum system to evolve between two distinguishable states, the quantum speed limit (QSL) characterizes the latent capability in speeding up of the system. It has found applications in determining the quantum superiority in many quantum technologies. However, previous results showed that such a speedup capability is generally destroyed by the environment induced decoherence in the Born-Markovian approximate dynamics. We here propose a scheme to recover the speedup capability in a dissipative continuous-variable system within the exact non-Markovian framework. It is found that the formation of a bound state in the energy spectrum of the total system consisting of the system and its environment can be used to restore the QSL to its noiseless performance. Giving an intrinsic mechanism in preserving the QSL, our scheme supplies a guideline to  speed up certain quantum tasks in practical continuous-variable systems.
\end{abstract}
\maketitle

\section{Introduction}

The quantum speed limit (QSL) quantifies the maximal speed at which a quantum system evolves under the constraint of quantum mechanics. Mandelstam and Tamm showed that, for a unitary dynamics governed by a Hamiltonian $\hat{H}$, the minimal evolution time between two orthogonal states is $\tau_{\text{MT}}=\pi\hbar/(2\Delta H)$ with $(\Delta H)^{2}\equiv\langle \hat{H}^{2}\rangle-\langle \hat{H}\rangle^{2}$ being the energy fluctuation~\cite{JPhys(USSR),PhysRevLett.65.1697}. This result provides a physical explanation to Heisenberg's energy-time uncertainty relation~\cite{JPhys(USSR),PhysRevA.31.2078,2007Timothy,2017Sebastian}. Later, Margolus and Levitin provided an alternative QSL time in terms of the energy difference $\tau_{\text{ML}}=\pi\hbar/(2\langle \hat{H}\rangle)$~\cite{MARGOLUS1998188,PhysRevA.82.022107}. Sun and Zheng derived a distinct QSL bound via the gauge invariant distance~\cite{PhysRevLett.123.180403}. The above three independent bounds are summarized in a unified form for both Hermitian and non-Hermitian quantum systems~\cite{PhysRevLett.127.100404}.

Recently, much effort has been devoted to generalizing the concept of QSL from closed systems to open systems~\cite{PhysRevLett.111.010402,PhysRevLett.110.050402,PhysRevLett.110.050403,PhysRevLett.115.210402,PhysRevLett.126.010601,2019Ken}. Deffner and Lutz derived a Margolus-Levitin-type bound on the minimal evolution time of open quantum systems~\cite{PhysRevLett.111.010402}. A generalized geometric interpretation for the Margolus-Levitin-like QSL was provided by Ref.~\cite{PhysRevX.6.021031}. From the application perspective, the QSL in open quantum systems is closely related to the greatest efficiency of charging power in quantum batteries~\cite{PhysRevLett.118.150601,PhysRevE.100.032107,PhysRevLett.125.040601}, the minimum operation time of quantum gates~\cite{PhysRevA.85.052327,PhysRevLett.126.220502}, the entropy production rate of nonequilibrium quantum thermodynamics~\cite{PhysRevLett.105.170402,PhysRevLett.113.260601,PhysRevLett.121.160602,PhysRevLett.121.070601,Nicholson2020,PhysRevX.11.011035}, as well as the quantum Fisher information in noisy quantum metrology~\cite{Giovannetti2011,PhysRevLett.112.120405,PhysRevX.6.021031,PhysRevLett.119.010403,PhysRevLett.125.120604,PhysRevX.10.021056}. Thus, how to establish a unified QSL bound, which is valid for both unitary and nonunitary evolutions, is of importance. Using the information geometric formalism is a possible solution~\cite{PhysRevX.6.021031,PhysRevLett.123.180403,PhysRevLett.127.100404,2017Sebastian,PhysRevLett.120.070401,Deffner2017}. Starting from a geometric perspective, Refs.~\cite{PhysRevLett.120.060409,Campaioli2019tightrobust} reported QSL bounds, which outperform the traditional bounds for both closed and open systems. As already shown in Refs.~\cite{PhysRevLett.123.180403,PhysRevLett.127.100404,PhysRevA.93.020105}, the QSL can be used to quantify the potential capability of speeding up for quantum systems. Such a speedup potency plays a leading role in quantum control~\cite{2017Sebastian,Aifer_2022}.

However, due to the decoherence induced by the inevitable system-environment interaction, the potency of quantum speedup generally vanishes in the Born-Markovian approximate decoherence dynamics ~\cite{PhysRevLett.110.050402,PhysRevLett.110.050403,PhysRevLett.115.210402,PhysRevLett.126.010601,PhysRevA.93.020105,PhysRevA.101.042107,PhysRevA.103.022221,PhysRevA.100.052305,2016Xu}. How to preserve such a capacity is of importance in the protocol of quantum technology and quantum control. On the other hand, most of the existing studies of the QSL in open quantum systems have focused on the discrete-variable case~\cite{PhysRevLett.111.010402,PhysRevLett.110.050402,PhysRevLett.110.050403,PhysRevLett.115.210402,PhysRevLett.126.010601,PhysRevA.93.020105,PhysRevA.100.052305,2016Xu,weiwu}. Very few studies concentrate on the continuous-variable case, especially in the non-Markovian dynamics. In this paper, we investigate the QSL in a dissipative continuous-variable system beyond the traditional paradigm of Born-Markovian approximation treatment. A bound-state based mechanism to realize a controllable QSL time in the noisy environment is revealed.

The paper is organized as follows. The QSL for a Gaussian continuous-variable system being applicable in both the closed and open systems is derived in Sec. \ref{secdef}. The non-Markovian decoherence effect on the QSL time is investigated in Sec. \ref{secnonmkv}. A mechanism to recover the ideal speedup capacity of the continuous-variable system under the non-Markovian noise is uncovered.  In Sec. \ref{seccomp}, we make a comparison of our scheme with the previous characterization schemes to the QSL in order to exhibit the universality of our result. Finally, a discussion and a summary are made in Sec. \ref{secdscs}.

\section{QSL in a Gaussian system}\label{secdef}

The QSL can be obtained from the viewpoint of the information geometry as follows. By introducing any kind of geodesic measure $\mathcal{L}=\mathcal{L}(\varrho_{\tau},\varrho_{0})$ quantifying the lower distance bound between two quantum states $\varrho_{\tau}$ and $\varrho_{0}$, an inequality is accordingly built as $\mathcal{L}\leq\ell\equiv\int\sqrt{d\ell^{2}}$. Here, $d\ell^{2}$ denotes the squared infinitesimal length between $\varrho_{\tau}$ and $\varrho_{\tau}+d\varrho_{\tau}$, which is regarded as the metric~\cite{Provost1980,2019Ken}, and thus $\ell$ is the length of the actual evolution path. Via introducing the time-averaged evolution speed $\bar{v}=\ell/\tau$, the QSL time is geometrically described as $\tau_{\mathrm{QSL}}\equiv\mathcal{L}/\bar{v}$, which implies $\tau_{\mathrm{QSL}}/\tau=\mathcal{L}/\ell$~\cite{PhysRevX.6.021031,2017Sebastian,2019Ken}. This result indicates that $\tau_{\mathrm{QSL}}/\tau$ characterizes the extent of the actual evolution path deviating from the geodesic path~\cite{PhysRevLett.123.180403,PhysRevLett.127.100404,PhysRevA.93.020105}. If $\tau_{\text{QSL}}/\tau=1$, the length of the actual evolution path saturates the geodesic one and there is no more space for speeding up. In contrary, the quantum system has a potential speedup capacity as long as $\tau_{\text{QSL}}/\tau<1$. The smaller the value of $\tau_{\mathrm{QSL}}/\tau$ is, the more speedup capability the system may possess. Therefore, $\tau_{\text{QSL}}/\tau$ is physically a characterization of the latent capability in speeding up of the quantum system. It has been found that such a capability has important applications in quantum technologies~\cite{PhysRevLett.118.150601,PhysRevE.100.032107,PhysRevLett.125.040601,PhysRevA.85.052327,PhysRevLett.126.220502,2017Sebastian,Aifer_2022}. It should be emphasized that the QSL bound considered in our paper is completely different from the so-called quantum brachistochrone problem~\cite{PhysRevLett.96.060503,PhysRevA.77.014103,PhysRevLett.111.260501}. The quantum brachistochrone problem commonly aims at designing an optimally controlled time-dependent Hamiltonian such that the shortest evolution time from a given initial state to a final one is achieved under a set of given constraints. It belongs to the research field of quantum optimal control. In this paper, we concentrate on an autonomous time-independent open system.

The Bures angle~\cite{doi:10.1080/09500349414552171,2017Sebastian,PhysRevLett.110.050402}
\begin{equation}
\mathcal{L}_{\text{B}}\equiv\text{arccos}\text{Tr}\bigg{(}\sqrt{\sqrt{\varrho_{0}}\varrho_{\tau}\sqrt{\varrho_{0}}}\bigg{)},
\end{equation}
is widely used to measure the geodesic length between $\varrho_{0}$ and $\varrho_{\tau}$. The corresponding metric known as the so-called Fisher-Rao metric relates to the famous quantum Fisher information as $d\ell^{2}=\frac{1}{4}\mathcal{F}_{\text{Q}}dt^{2}$ ~\cite{PhysRevX.6.021031,2017Sebastian,PhysRevLett.110.050402,PhysRevLett.72.3439}. Here, the quantum Fisher information $\mathcal{F}_{\text{Q}}$ is defined by $\mathcal{F}_{\text{Q}}=\text{Tr}(\varrho_{t}\hat{L}^{2})$ with $\hat{L}$ determined by $\dot{\varrho}_{t}=(\hat{L}\varrho_{t}+\varrho_{t}\hat{L})/2$. Then, the averaged speed $\bar{v}$ and the QSL time $\tau_{\mathrm{QSL}}$ are derived as~\cite{PhysRevX.6.021031,2017Sebastian,PhysRevLett.110.050402,PhysRevA.103.022210,H_rnedal_2022}
\begin{eqnarray}
\bar{v}&=&\frac{1}{2\tau}\int_{0}^{\tau}dt\sqrt{\mathcal{F}_{\text{Q}}},\\
\tau_{\mathrm{QSL}}&=&\frac{\mathcal{L}_{\text{B}}}{\bar{v}}=\frac{\mathcal{L}_{\text{B}}}{\ell}\tau.\label{eq:eq3}
\end{eqnarray}
It is found that $\mathcal{L}_{\text{B}}$ and $\mathcal{F}_{\text{Q}}$ naturally reduce to $\pi/2$ and $4(\langle\psi|\hat{H}^{2}|\psi\rangle-\langle\psi|\hat{H}|\psi\rangle^{2})$, respectively, in the special case of the pure states under the unitary evolution, i.e., $\varrho_{0}=|\psi\rangle\langle\psi|$ and $\varrho_{\tau}=|\psi_{\perp}\rangle\langle\psi_{\perp}|$ with $\langle\psi|\psi_{\perp}\rangle=0$. Then, $\tau_{\text{QSL}}$ recovers the well-known Mandelstam-Tamm bound.

Here, we consider a single-mode continuous-variable system consisting of a pair of annihilation and creation operators $\hat{A}=\{\hat{a},\hat{a}^{\dagger}\}$. The characteristic function of the system is defined as $\chi(\pmb{\xi})\equiv\text{Tr}[\varrho_{t}\mathcal{\hat{D}}(\pmb{\xi})]$~\cite{RevModPhys.77.513,_afr_nek_2015}, where $\mathcal{\hat{D}}(\pmb{\xi})=\exp(\hat{A}^{\dagger}K\pmb{\xi})$, with $K=\text{Diag}(1,-1)$ and $\pmb{\xi}=(\xi,\xi^{*})^{\text{T}}$, is the Weyl displacement operator. If the characteristic function has a Gaussian form $\chi(\pmb{\xi})=\exp(-\frac{1}{4}\pmb{\xi}^{\dagger}\pmb{\sigma}\pmb{\xi}-i\pmb{d}^{\dagger}K\pmb{\xi})$, then such a continuous-variable system is called a Gaussian system. Its characteristic function is fully determined by the displacement vector $\pmb{d}_{t}$ with $d_{t}^{j}=\text{Tr}(\varrho_{t}\hat{A}_{j})$ and the covariance matrix $\pmb{\sigma}_{t}$ with $\sigma_{t}^{ij}=\text{Tr}(\varrho_{t}\{\hat{A}_{i}-d_{i},\hat{A}^{\dagger}_{j}-d^{*}_{j}\})$. For a Gaussian continuous-variable system, the Bures angle reads $\mathcal{L}_{\text{B}}=\text{arccos}\sqrt{F}$. Here, $F$ is the quantum fidelity and is calculated by using the displacement vectors
and the covariance matrices as~\cite{Scutaru_1998,_afr_nek_2015,PhysRevA.86.022340}
\begin{equation}
F=\frac{2\exp[-(\pmb{d}_{0}-\pmb{d}_{t})^{\dagger}(\pmb{\sigma}_{0}+\pmb{\sigma}_{t})^{-1}(\pmb{d}_{0}-\pmb{d}_{t})]}{\sqrt{\Gamma}+\sqrt{\Lambda}-\sqrt{(\sqrt{\Gamma}+\sqrt{\Lambda})^{2}-\Pi}},
\end{equation}
where $\Pi=\text{Det}(\pmb{\sigma}_{0}+\pmb{\sigma}_{t})$, $\Gamma=\text{Det}(\mathbf{1}+K\pmb{\sigma}_{0}K\pmb{\sigma}_{t})$, and $\Lambda=\text{Det}(\pmb{\sigma}_{0}+K)\text{Det}(\pmb{\sigma}_{t}+K)$. On the other hand, the quantum Fisher information with respect to the evolution time for a Gaussian system is calculated by~\cite{_afr_nek_2018}
\begin{equation}
\mathcal{F}_{\text{Q}}=\frac{1}{2}\text{Vec}[\dot{\pmb{\sigma}}_{t}]^{\dagger}M^{-1}\text{Vec}[\dot{\pmb{\sigma}}_{t}]+2\dot{\pmb{d}}_{t}^{\dagger}\pmb{\sigma}^{-1}_{t}\dot{\pmb{d}}_{t},
\end{equation}
where $\text{Vec}[\cdot]$ denotes the vectorization of a given matrix and $M=\pmb{\sigma}_{t}^{*}\otimes\pmb{\sigma}_{t}-K\otimes K$. From these results, as long as $\pmb{d}_{t}$ and $\pmb{\sigma}_{t}$ are known, the QSL in a Gaussian continuous-variable system is fully determined.

Let us first consider the QSL of a quantum harmonic oscillator in the ideal case of a unitary evolution governed by $\hat{H}_{\text{s}}=\omega_{0}\hat{a}^{\dagger}\hat{a}$. The initial state is chosen as a coherent state, namely, $\varrho_{\text{s}}(0)=\mathcal{\hat{D}}(\alpha)|0\rangle\langle 0|\mathcal{\hat{D}}(\alpha)^{\dagger}$. It is readily derived that $\pmb{d}_{t}=(\alpha e^{-i\omega_{0}t},\alpha^{*}e^{i\omega_{0}t})^{\text{T}}$ and $\pmb{\sigma}_{t}=\mathbf{1}$, which lead to $F=e^{-2|\alpha|^{2}[1-\cos(\omega_{0}\tau)]}$ and $\mathcal{F}_{\text{Q}}=4|\alpha|^{2}\omega_{0}^{2}$. Thus, we have $\bar{v}=|\alpha|\omega_{0}$, which is a time-independent constant, and thus
\begin{equation}\label{eq:eq4}
{\tau_{\mathrm{QSL}}\over\tau}=\frac{\arccos e^{-|\alpha|^{2}[1-\cos(\omega_{0}\tau)]}}{|\alpha|\omega_{0}\tau}.
\end{equation}
Equation \eqref{eq:eq4} reveals that the QSL time behaves as $\tau_{\mathrm{QSL}}/\tau\propto\tau^{-1}$ with the actual evolution time $\tau$. It means $\tau_{\mathrm{QSL}}/\tau$ approaches to zero in the large-$\tau$ regime. Such a result implies that the harmonic oscillator has an infinite speedup capability in this noiseless case.

\section{QSL in a noisy environment}\label{secnonmkv}

Next, we consider a more practical situation in which the harmonic oscillator is coupled to a dissipative bosonic environment and experiences a decoherence. The Hamiltonian of the total system reads
\begin{equation}\label{eq:eq5}
\begin{split}
\hat{H}=&\hat{H}_{\text{s}}+\sum_{k}\omega_{k}\hat{b}_{k}^{\dagger}\hat{b}_{k}+\sum_{k}\Big{(}g_{k}\hat{a}^{\dagger}\hat{b}_{k}+\mathrm{H}.\mathrm{c}.\Big{)},
\end{split}
\end{equation}
where $\hat{b}_{k}$ denotes the annihilation operator of the $k$th environmental mode with frequency $\omega_{k}$, and the parameter $g_{k}$ is the coupling strength between the harmonic oscillator and the $k$th environmental mode. The coupling strength is further characterized by the spectral density $J(\omega)\equiv\sum_{k}|g_{k}|^{2}\delta(\omega-\omega_{k})$. We consider that $J(\omega)$ explicitly takes the following Ohmic-family form:
\begin{equation}\label{eq:eq6}
J(\omega)=\eta\omega^{s}\omega_{c}^{1-s}e^{-\omega/\omega_{c}},
\end{equation}
where $\eta$ is a dimensionless coupling constant, $\omega_{c}$ is a cutoff frequency, and $s$ is the so-called Ohmicity parameter. Depending on the value of $s$, the environment can be classified into the sub-Ohmic for $0<s<1$, the Ohmic for $s=1$, and the super-Ohmic for $s>1$.

Considering the environment is initially prepared in its vacuum state and using Feynman-Vernon's influence functional method to partially trace out the degrees of freedom of the dissipative environment, we obtain an exact non-Markovian master equation for the harmonic oscillator as~\cite{PhysRevA.76.042127,PhysRevE.90.022122,PhysRevA.103.L010601}
\begin{eqnarray}\label{eq:eq7}
\dot{\varrho}_{t}=-i\big{[}\Omega(t)\hat{a}^{\dagger}\hat{a},\varrho_{t}\big{]}+\gamma(t)\Big{(}2\hat{a}\varrho_{t}\hat{a}^\dag-\{\hat{a}^\dag\hat{a},\varrho_{t}\}\Big{)},
\end{eqnarray}
where $\Omega(t)=-\text{Im}[\dot{u}(t)/u(t)]$ is the renormalized frequency and $\gamma(t)=-\text{Re}[\dot{u}(t)/u(t)]$ is the decay rate induced by the dissipative environment. The time-dependent coefficient $u(t)$ is determined by
\begin{equation}\label{eq:eq8}
\dot{u}(t)+i\omega_{0}u(t)+\int_{0}^{t}d\tau\mu(t-\tau)u(\tau)=0,
\end{equation}
with $u(0)=1$ and $\mu(t)\equiv\int_{0}^{\infty}d\omega J(\omega)e^{-i\omega t}$.

In order to compare with that of the noiseless ideal case, we still choose that the quantum harmonic oscillator is initially in a coherent state. Then, solving the master equation \eqref{eq:eq7}, we calculate the exact expressions of the displacement vector and the covariance matrix as $\pmb{d}_{t}=[\alpha u(t),\alpha^{*} u(t)^{*}]^{\mathrm{T}}$ and $\pmb{\sigma}_{t}=\mathbf{1}$. With the above expressions at hand, the time-averaged speed $\bar{v}$ and the geodesic distance $\mathcal{L}_{\text{B}}$ in the noise case are straightforwardly computed:
\begin{eqnarray}
\bar{v}&=&\frac{1}{\tau}\int_{0}^{\tau}dt|\alpha\dot{u}(t)|,\label{ourdv}\\
\mathcal{L}_{\text{B}}&=&\arccos \sqrt{e^{-|\alpha[1-u(\tau)]|^{2}}}.
\end{eqnarray}
We consider the QSL of the system relaxing to its steady state by choosing $\tau$ sufficiently large. The QSL time derived under such a condition reflects the equilibration efficiency of the dissipative harmonic oscillator. The controllability of this equilibration efficiency is vital in suppressing the detrimental effect of the decoherence in practical quantum technologies.

\begin{figure}
\centering
\includegraphics[angle=0,width=8.75cm]{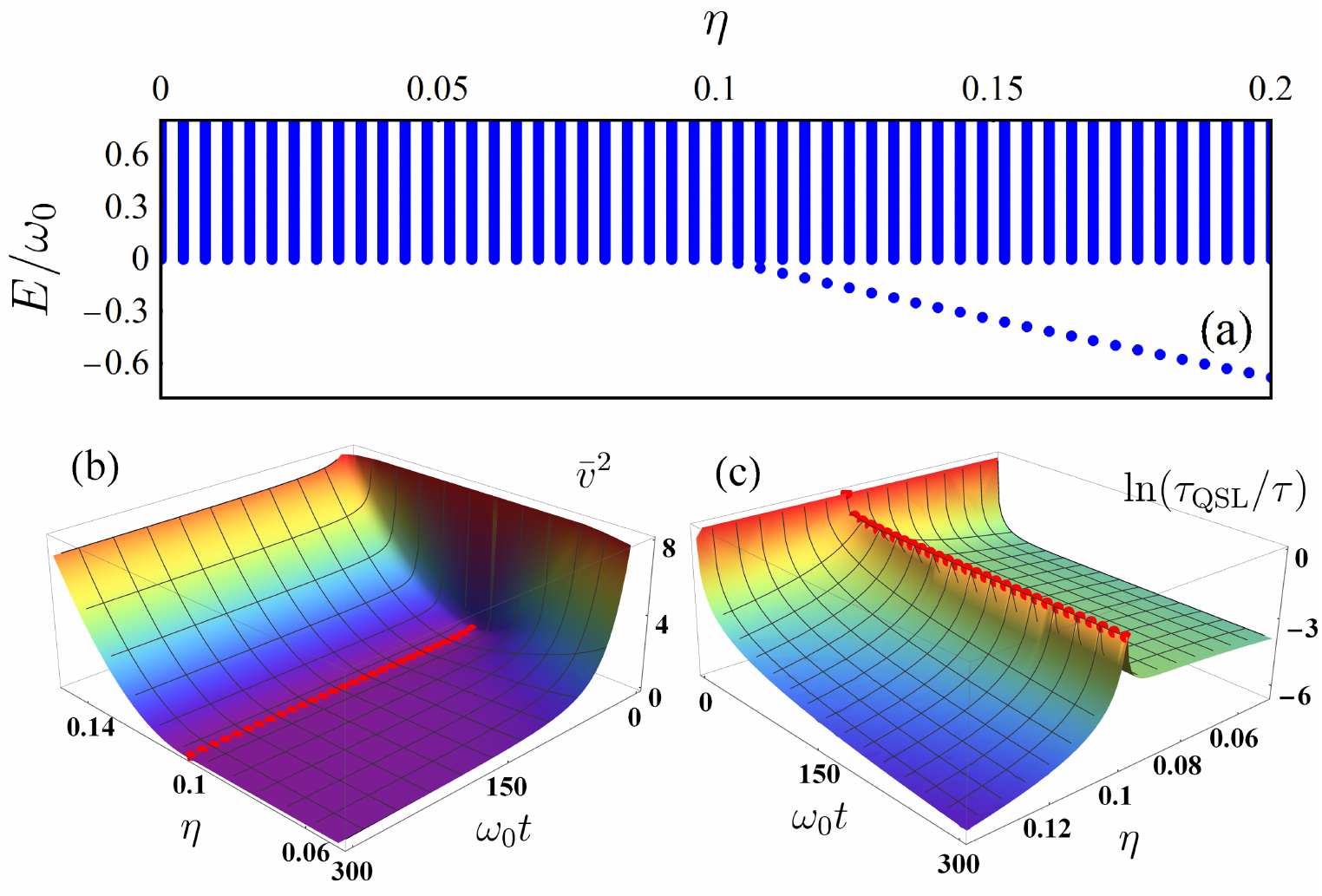}
\caption{(a) Energy spectrum of $\hat{H}$ in the single-excitation subspace. Exact non-Markovian result of the average speed (b) and the QSL time (c) with different $\eta$ and $\omega_{0}t$. The red circles highlight the positions of the threshold coupling strength $\eta=0.1$ at which the bound state occurs. Other parameters are $|\alpha|=10$, $s=1$, and $\omega_{c}=10\omega_{0}$.}\label{fig:fig1}
\end{figure}
If the system-environment coupling is weak and the characteristic time scale of the environmental correlation function is much smaller than that of the system, one can safely apply the Born-Markovian approximation to Eq.~(\ref{eq:eq8}). Under such a circumstance, one calculates~\cite{PhysRevE.90.022122,PhysRevA.103.L010601} $u^\text{MA}(t)\simeq e^{-\{\kappa+i[\omega_{0}+\Delta(\omega_{0})]\}t}$, where $\kappa=\pi J(\omega_{0})$ is the Markovian decay coefficient and $\Delta(\omega_{0})=\mathcal{P}\int _0^\infty d\omega{J(\omega)\over \omega_{0}-\omega}$ is the environmentally induced frequency shift. With the approximate expression of $u^\text{MA}(t)$ at hand, we find, under the Born-Markovian approximation, $\bar{v}=|\alpha|\sqrt{\kappa^{2}+\omega_{0}^{2}}(1-e^{-\kappa\tau})/(\kappa\tau)$ and
\begin{equation}\label{eq:eq9}
\lim_{\tau\rightarrow\infty}\frac{\tau_{\mathrm{QSL}}}{\tau}=\frac{\kappa\arccos e^{-\frac{1}{2}|\alpha|^{2}}}{|\alpha|\sqrt{\kappa^{2}+\omega_{0}^{2}}},
\end{equation}
where we have dropped the contributions from the frequency shift term $\Delta(\omega_{0})$. Equation \eqref{eq:eq9} reduces to the one of the noiseless case in the limit $\kappa\rightarrow0$. We here choose a large $\tau$, which means we focus on the QSL time from the initial coherent state to the equilibrium state. In this limit, we find that $\bar{v}$ reduces to zero and $\tau_{\mathrm{QSL}}/\tau$ evolves to a time-independent value. This result implies that the speedup potency of the system is destroyed by the Born-Markovian decoherence. A similar result was also reported in several previous references~\cite{PhysRevA.101.042107,PhysRevA.103.022221,Deffner2017,PRXQuantum.2.040349}.

Going beyond the Born-Markovian approximation, the results of $\bar{v}$ and $\tau_{\text{QSL}}/\tau$ are obtainable by numerically solving Eq.~(\ref{eq:eq8}). However, via analyzing the long-time behavior of $u(\tau)$, we calculate their analytically asymptotic forms in the limit $\tau\rightarrow\infty$, which shall help us to build up a more clear physical picture on our results. To this aim, we apply a Laplace transform to $u(\tau)$ and find
$\tilde{u}(z)\equiv\int_{0}^{\infty}d\tau u(\tau)e^{-z\tau}=[z+i\omega_{0}+\int_0^\infty d\omega{J(\omega)\over z+i\omega}]^{-1}$. The solution of $u(\tau)$ is immediately obtained by applying an inverse Laplace
transform to $\tilde{u}(z)$, which is exactly done by finding the poles of $\tilde{u}(z)$ from the following transcendental equation:
\begin{equation}\label{eq:eq10}
y(\varpi)\equiv\omega_{0}-\int_0^\infty d\omega{J(\omega)\over\omega-\varpi} =\varpi,
\end{equation}
with $\varpi=iz$. It is necessary to point out that the root of the above equation is just the eigenenergy of $\hat{H}$ in the single-excitation subspace. To be more specific, we express the single-excitation eigenstate as $|\Psi\rangle=(x\hat{a}^{\dagger}+\sum_{k}y_{k}\hat{b}_{k}^{\dagger})|0,\{0_k\}\rangle$. Substituting it into $\hat{H}|\Psi\rangle=E|\Psi\rangle$, one finds the energy eigen-equation as $E-\omega_{0}-\sum_{k}g_{k}^{2}/(\omega_{k}-E)=0$, which retrieves Eq.~(\ref{eq:eq10}) via simply replacing $E$ by $\varpi$. This result implies that, although the subspaces with arbitrary excitation number are involved in the reduced dynamics, the dynamics of $u(\tau)$ is essentially determined by the single-excitation energy spectrum characteristic of $\hat{H}$. Because $y(\varpi)$ is a monotonically decreasing function in the regime $\varpi<0$, Eq.~(\ref{eq:eq10}) potentially has one isolated root $E_b$ in this regime provided $y(0)<0$. While $y(\varpi)$ is not well analytic in the regime $\varpi>0$, Eq.~(\ref{eq:eq10}) has infinite roots in this regime and forms a continuous energy band. We call the eigenstate corresponding to the isolated eigenenergy $E_b$ the bound state. Then, after applying the inverse Laplace transform and using the residue theorem, we obtain~\cite{PhysRevE.90.022122,PhysRevA.103.L010601}
\begin{equation*}
u(\tau)=Ze^{-iE_b \tau}+\int_{0}^{\infty}\frac{J(\omega)e^{-i\omega \tau}d\omega}{[\omega-\omega_{0}-\Delta(\omega)]^{2}+[2\pi J(\omega)]^2},
\end{equation*}
where the first term with $Z\equiv[1+\int_0^\infty{J(\omega)d\omega\over(E_b-\omega)^2}]^{-1}$ is contributed from the potentially formed bound state energy $E_{b}$, and the second term is from the band energy which approaches to zero in the long-time regime due to out-of-phase interference. Thus, if the bound state is absent, then we have $u(\infty)=0$, which leads to a complete decoherence; while if the bound state with energy $E_{b}$ is formed, then we have $u(\infty)\simeq Ze^{-iE_b\tau}$, which implies a dissipationless dynamics. The condition of forming the bound state for the Ohmic-family spectral density is evaluated via $y(0)<0$ as $\omega_{0}-\eta\omega_c\Gamma(s)< 0$, where $\Gamma(s)$ is Euler's gamma function.

\begin{figure}
\centering
\includegraphics[angle=0,width=8.85cm]{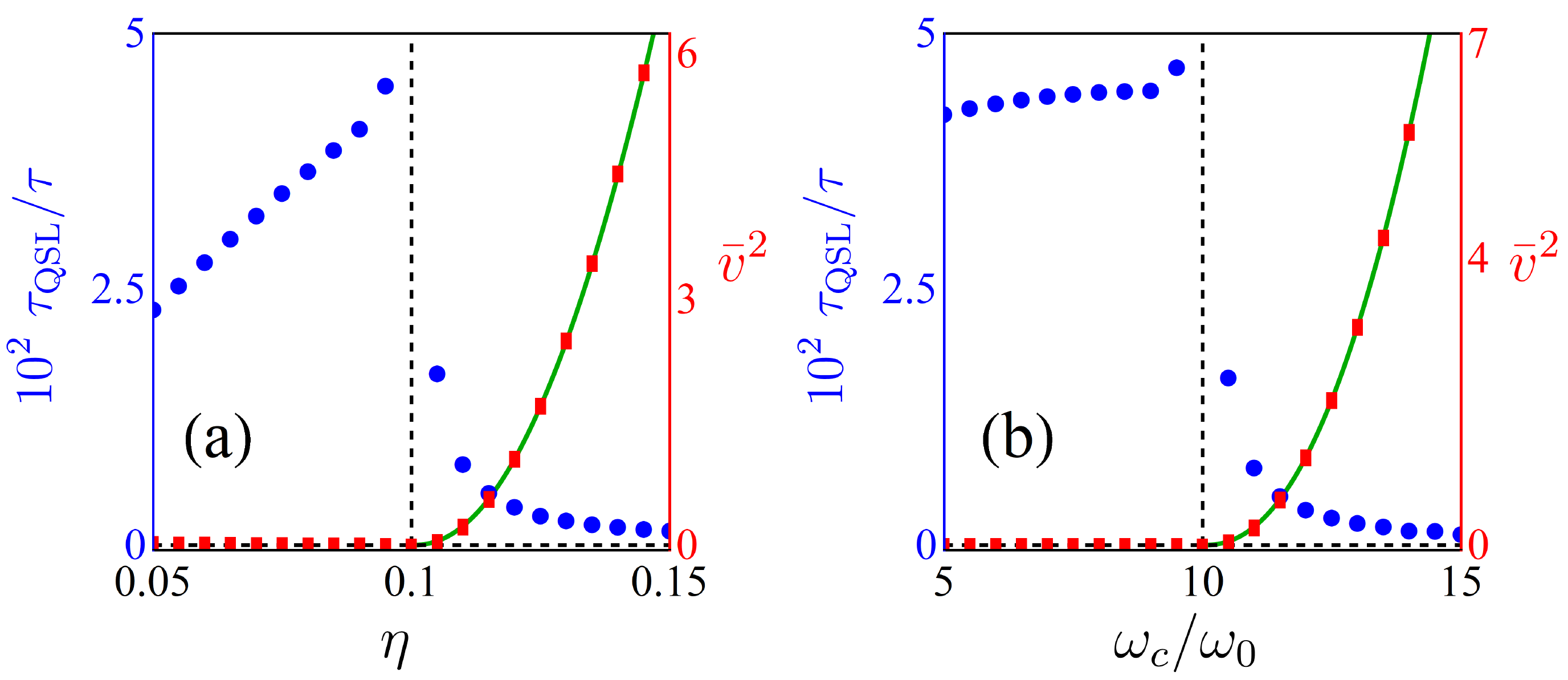}
\caption{(a) Steady-state average speed (red rectangles) and QSL time (blue circles) as a function of $\eta$ (a) and $\omega_{c}/\omega_{0}$ (b) when $\omega_{0}\tau=400$. The green solid lines are obtained from the analytical result $\bar{v}^{2}\simeq|\alpha|^{2}Z^{2}E_{b}^{2}$. Other parameters are chosen as $|\alpha|=10$ and $s=1$.}\label{fig:fig2}
\end{figure}

In the absence of the bound state, it is natural to expect a consistent result with that under the Born-Markovian approximation because $u(\tau)$ approaches zero eventually. In contrary, with the long-time expression of $u(\infty)\simeq Ze^{-iE_b\tau}$ in the presence of the bound state, we find $\bar{v}\simeq|\alpha ZE_{b}|$ and
\begin{equation}\label{eq:eq12}
\frac{\tau_{\mathrm{QSL}}}{\tau}=\frac{\arccos e^{-\frac{1}{2}|\alpha|^{2}[1+Z^{2}-2Z\cos(E_{b}\tau)]}}{|\alpha ZE_{b}|\tau}.
\end{equation}
We see from Eq. \eqref{eq:eq12} that, in the limit $\tau\rightarrow\infty$, $\bar{v}$ approaches to a non-zero value, while $\tau_{\mathrm{QSL}}/\tau$ reduces to zero in the form of $\tau^{-1}$. These results are verified by exact numerical simulations (see Fig.~\ref{fig:fig1}) and are completely different from those under the Born-Markovian case~\cite{PhysRevA.101.042107,PhysRevA.103.022221,Deffner2017}. Compared to that of the noiseless ideal case, recovering the relation $\tau_{\mathrm{QSL}}/\tau\propto\tau^{-1}$ means the potency of quantum speedup is fully retrieved. In Fig.~\ref{fig:fig2}, we plot the long-time steady-state $\bar{v}$ and $\tau_{\text{QSL}}/\tau$ as functions of $\eta$ and $\omega_{c}/\omega_{0}$. It confirms that there exists a threshold from no-speedup to speedup regimes matching well with the position of forming the bound state. Our result implies that the time-averaged quantum speed and the QSL time are controllable via engineering the energy spectrum of the whole oscillator-environment system.

\section{Comparisons with previous studies}\label{seccomp}

In several previous articles~\cite{PhysRevA.101.042107,PhysRevA.103.022221,Deffner2017}, the Wigner function and the Wasserstein distance are used to calculate the QSL time in Gaussian continuous-variable systems. As displayed in Refs.~\cite{PhysRevA.101.042107,PhysRevA.103.022221,Deffner2017}, within the framework of the Wigner representation, the geodesic length between two Wigner distributions, $W_{\tau}(\pmb\zeta)$ and $W_{0}(\pmb\zeta)$ with $\pmb\zeta=(x,p)^{\text{T}}=(\frac{\xi+\xi^{*}}{\sqrt{2}},\frac{\xi-\xi^{*}}{\sqrt{2}i})^{\text{T}}$ being the quadrature vector, is quantified by using the Wasserstein-$2$ distance as
\begin{equation}
\begin{split}
\mathcal{L}_{W}=&\|W_{\tau}(\pmb\zeta)-W_{0}(\pmb\zeta)\|_{2}\\
\equiv&\big{\{}\int d\pmb\zeta|W_{\tau}(\pmb\zeta)-W_{0}(\pmb\zeta)|^{2}\big{\}}^{\frac{1}{2}}.
\end{split}
\end{equation}
Then the averaged evolution speed $\bar{v}_{W}$ and the QSL time $\tau^{W}_{\mathrm{QSL}}$ in the Wigner space are established as~\cite{Deffner2017}
\begin{eqnarray}
\bar{v}_{W}&=&\frac{1}{\tau}\int_{0}^{\tau}dt\|\dot{W}_{t}(\pmb\zeta)\|_{2},\\
\tau^{W}_{\mathrm{QSL}}&=&\frac{\mathcal{L}_{W}}{\bar{v}_{W}}.
\end{eqnarray}
The above formalism was further generalized to the Wasserstein-$p$-distance cases with $p=1,2$ and $+\infty$, but computing these Wasserstein distances is rather complicated~\cite{Deffner2017}.

\begin{figure}
\centering
\includegraphics[angle=0,width=8.75cm]{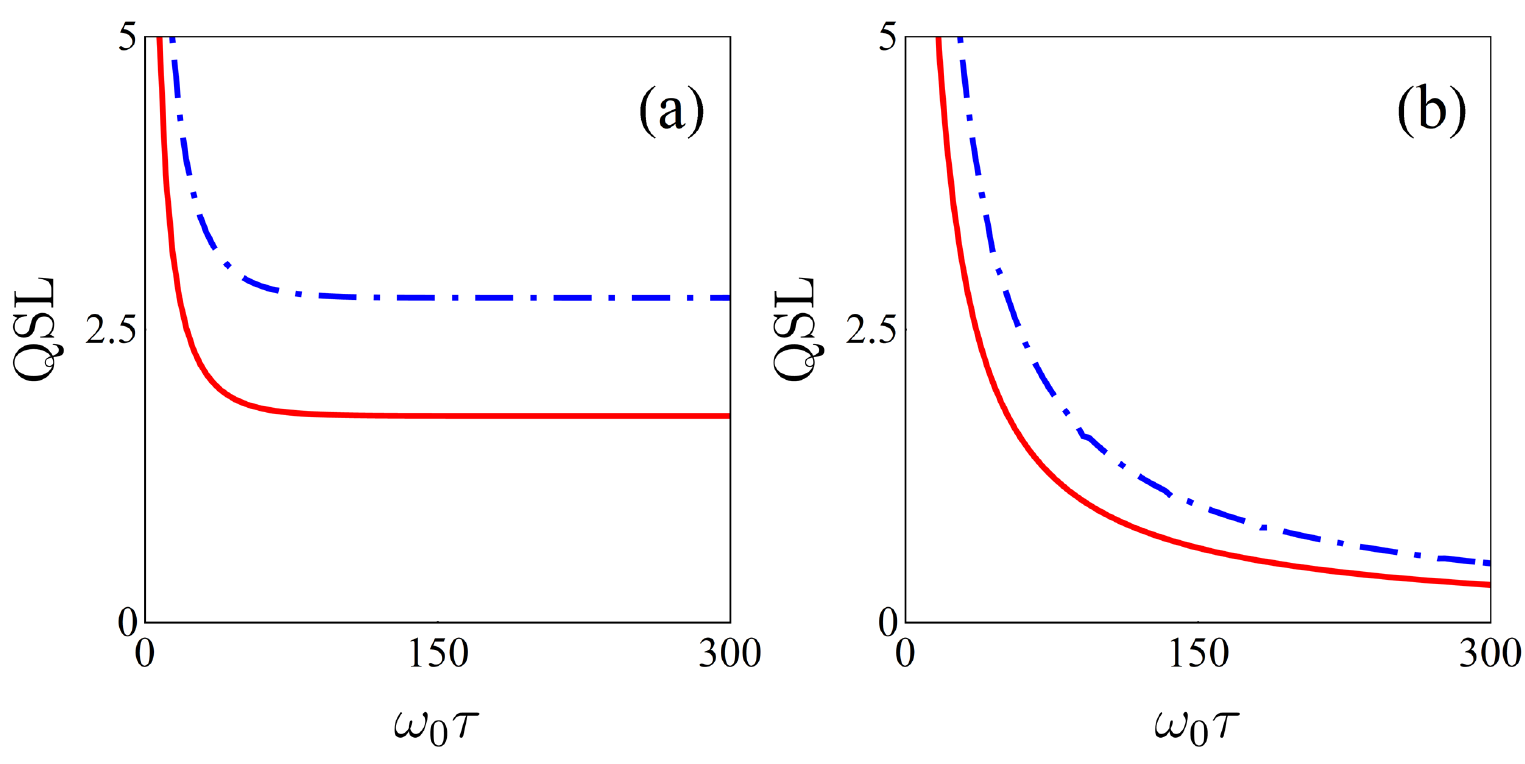}
\caption{(a) The QSL time $10^{2}\tau_{\text{QSL}}/\tau$ (blue dot-dashed line) and $10^{2}\tau^{W}_{\text{QSL}}/\tau$ (red solid line) are plotted as a function of $\omega_{0}\tau$ in the absence of a bound state when $\eta=0.06$. (b) The same as (a), but in the presence of a bound state with $\eta=0.12$. Other parameters are chosen as $|\alpha|=10$ and $s=1$.}\label{fig:fig3}
\end{figure}

For our dissipative harmonic oscillator system, the exact expression of the Wigner function is given by~\cite{Chen_2019,PhysRevA.103.L010601}
\begin{equation}
W_{t}(\pmb\zeta)=\frac{e^{-\frac{1}{2}(\pmb{\zeta}-\tilde{\pmb{d}}_{t})^{\mathrm{T}}\tilde{\pmb{\sigma}}_{t}^{-1}(\pmb{\zeta}-\tilde{\pmb{d}}_{t})}}{\pi\sqrt{|\mathrm{Det}\tilde{\pmb{\sigma}}_{t}|}},
\end{equation}
where $\tilde{\pmb{d}}_{t}=(\sqrt{2}\text{Re}[\alpha u(t)],\sqrt{2}\text{Im}[\alpha u(t)])^{\text{T}}$ and $\tilde{\pmb{\sigma}}_{t}=\frac{1}{2}\pmb{\sigma}_{t}$. With the above expression at hand, we find
\begin{eqnarray}
\bar{v}_{W}&=&\frac{2}{\sqrt{\pi}\tau}\int_{0}^{\tau}dt|\alpha\dot{u}(t)|,\label{ssdfdd}\\
\mathcal{L}_{W}&=&\frac{2}{\sqrt{\pi}}\big{\{}1-e^{-2|\alpha|^{2}([\text{Re}u(\tau)-1]^{2}+[\text{Im}u(\tau)-1]^{2})}\big{\}}^{\frac{1}{2}}.
\end{eqnarray}
It is immediately observed that Eq. \eqref{ssdfdd} matches Eq. \eqref{ourdv} except for a trivial pre-factor $2/\sqrt{\pi}$. Moreover, in the limit $\tau\rightarrow\infty$, we find that $\tau_{\text{QSL}}^{W}/\tau$ approaches to a non-zero constant in the absence of the bound state, and $\tau_{\text{QSL}}^{W}/\tau\propto\tau^{-1}$ in the presence of the bound state (see Fig.~\ref{fig:fig3}). This conclusion is completely consistent with that of $\tau_{\text{QSL}}$ obtained in Sec. \ref{secnonmkv}. It demonstrates that our bound-state-based QSL-controlling scheme is universal to different definitions of QSL time.

Next, we compare the tightness of $\tau_{\text{QSL}}$ and $\tau_{\text{QSL}}^{W}$. Based on our numerical simulations, $\tau_{\text{QSL}}$ is not always tighter than $\tau_{\text{QSL}}^{W}$ during the whole relaxation process. However, as displayed in Fig.~\ref{fig:fig3}, the value of $\tau_{\text{QSL}}$ is larger than $\tau_{\text{QSL}}^{W}$ in the large-$\tau$ regime irrespective of whether the bound state is formed or not. These results mean that the QSL considered in this paper can be tighter than the previous one derived by using the Wigner function and Wasserstein-2 distance. Furthermore, it is noted that, although both our paper and the ones in Refs.~\cite{PhysRevA.101.042107,PhysRevA.103.022221,Deffner2017} provide a computable way to obtain the QSL time in Gaussian continuous-variable systems, the QSL formulation in our paper is strictly established in the differential geometry, which is more rigorous in the mathematical sense. In fact, the Fisher-Rao metric employed by us is a contractive Riemannian metric on the set of density operators. As discussed in Ref.~\cite{PhysRevX.6.021031}, such a peculiar mathematical property may help us to find the tightest bound on the QSL time.

\section{Discussion and Summary}\label{secdscs}

It is necessary to emphasize that our bound-state based QSL-controlling scheme is independent of the choice of the spectral density. Although only the Ohmic-family form is considered in this paper, our result is straightforwardly generalizable to other cases without difficulties. The bound-state effect, which generally appears in the non-Markovian regime, is the crucial ingredient in our scheme of achieving a steerable QSL. How to generate the bound state is the main point in realizing our control scheme from an experimental perspective. Fortunately, thanks to the rapid development in the state-of-the-art technique of quantum optics experiments, the bound state and its dynamical effect have been observed in circuit quantum electrodynamics architecture~\cite{Liu2017} and matter-wave systems~\cite{Krinner2018}. With the help of the reservoir engineering technique, the primary parameters in the spectral density $J(\omega)$ are experimentally controllable. The spectral density of a quantum emitter acting as an open system coupling to the surface-plasmon polariton as an environment is adjustable by changing the distance between them \cite{Andersen2011,PhysRevB.95.161408}. For a reservoir consisting of ultracold atomic gas, the Ohmicity parameter $s$ is tunable from the sub-Ohmic to the super-Ohmic forms by increasing the scattering length of the gas via Feshbach resonances~\cite{PhysRevA.84.031602}. These experimental achievements provide a strong support to our theoretical investigations. As a final remark, our present paper is completely different from Ref.~\cite{weiwu}. We here investigate the QSL time derived by the Fisher-Rao metric of a continuous-variable system. However, Ref.~\cite{weiwu} considered the Fubini-Study-metric-based QSL of a dissipative two-level system. From the technical point of view, we need an exact expression of the deterministic quantum master equation to obtain the QSL. In contrast, Ref.~\cite{weiwu} proposed a scheme to calculate the QSL time via solving the stochastic Schr\"{o}dinger equation governed by an effective non-Hermitian Hamiltonian. Thus, neither the conclusions nor the methodology of Ref.~\cite{weiwu} is directly applicable to our present paper.

In summary, by making use of an exact non-Markovian treatment, we investigate the time-averaged evolution speed and the QSL time in an open continuous-variable quantum system. It is revealed that the formation of a bound state in the energy spectrum of the whole system-environment system in the single-excitation subspace is beneficial for recovering the speedup potency of an open system, which is generally destroyed under the Born-Markovian approximation. Compared with the previous studies~\cite{PhysRevA.101.042107,PhysRevA.103.022221,Deffner2017}, our result provides a tighter QSL time in the dissipative continuous-variable quantum systems. Being experimentally realizable in realistic platforms, our bound-state based QSL-controlling scheme opens an avenue to control the QSL of open system via engineering the energy-spectrum characteristic of the total system consisting of the open system and its environment.

\section*{Acknowledgments}

The work is supported by the National Natural Science Foundation of China (Grants No. 11875150, No. 12275109, No. 11834005, and No. 12047501).
	
\bibliography{reference}

\begin{thebibliography}{66}%
\makeatletter
\providecommand \@ifxundefined [1]{%
 \@ifx{#1\undefined}
}%
\providecommand \@ifnum [1]{%
 \ifnum #1\expandafter \@firstoftwo
 \else \expandafter \@secondoftwo
 \fi
}%
\providecommand \@ifx [1]{%
 \ifx #1\expandafter \@firstoftwo
 \else \expandafter \@secondoftwo
 \fi
}%
\providecommand \natexlab [1]{#1}%
\providecommand \enquote  [1]{``#1''}%
\providecommand \bibnamefont  [1]{#1}%
\providecommand \bibfnamefont [1]{#1}%
\providecommand \citenamefont [1]{#1}%
\providecommand \href@noop [0]{\@secondoftwo}%
\providecommand \href [0]{\begingroup \@sanitize@url \@href}%
\providecommand \@href[1]{\@@startlink{#1}\@@href}%
\providecommand \@@href[1]{\endgroup#1\@@endlink}%
\providecommand \@sanitize@url [0]{\catcode `\\12\catcode `\$12\catcode
  `\&12\catcode `\#12\catcode `\^12\catcode `\_12\catcode `\%12\relax}%
\providecommand \@@startlink[1]{}%
\providecommand \@@endlink[0]{}%
\providecommand \url  [0]{\begingroup\@sanitize@url \@url }%
\providecommand \@url [1]{\endgroup\@href {#1}{\urlprefix }}%
\providecommand \urlprefix  [0]{URL }%
\providecommand \Eprint [0]{\href }%
\providecommand \doibase [0]{http://dx.doi.org/}%
\providecommand \selectlanguage [0]{\@gobble}%
\providecommand \bibinfo  [0]{\@secondoftwo}%
\providecommand \bibfield  [0]{\@secondoftwo}%
\providecommand \translation [1]{[#1]}%
\providecommand \BibitemOpen [0]{}%
\providecommand \bibitemStop [0]{}%
\providecommand \bibitemNoStop [0]{.\EOS\space}%
\providecommand \EOS [0]{\spacefactor3000\relax}%
\providecommand \BibitemShut  [1]{\csname bibitem#1\endcsname}%
\let\auto@bib@innerbib\@empty
\bibitem [{\citenamefont {Mandelstam}\ and\ \citenamefont
  {Tamm}(1945)}]{JPhys(USSR)}%
  \BibitemOpen
  \bibfield  {author} {\bibinfo {author} {\bibfnamefont {L.}~\bibnamefont
  {Mandelstam}}\ and\ \bibinfo {author} {\bibfnamefont {I.}~\bibnamefont
  {Tamm}},\ }\bibfield  {title} {\enquote {\bibinfo {title} {The uncertainty
  relation between energy and time in nonrelativistic quantum mechanics},}\
  }\href@noop {} {\bibfield  {journal} {\bibinfo  {journal} {J. Phys.}\
  }\textbf {\bibinfo {volume} {9}},\ \bibinfo {pages} {249} (\bibinfo {year}
  {1945})}\BibitemShut {NoStop}%
\bibitem [{\citenamefont {Anandan}\ and\ \citenamefont
  {Aharonov}(1990)}]{PhysRevLett.65.1697}%
  \BibitemOpen
  \bibfield  {author} {\bibinfo {author} {\bibfnamefont {J.}~\bibnamefont
  {Anandan}}\ and\ \bibinfo {author} {\bibfnamefont {Y.}~\bibnamefont
  {Aharonov}},\ }\bibfield  {title} {\enquote {\bibinfo {title} {Geometry of
  quantum evolution},}\ }\href {\doibase 10.1103/PhysRevLett.65.1697}
  {\bibfield  {journal} {\bibinfo  {journal} {Phys. Rev. Lett.}\ }\textbf
  {\bibinfo {volume} {65}},\ \bibinfo {pages} {1697--1700} (\bibinfo {year}
  {1990})}\BibitemShut {NoStop}%
\bibitem [{\citenamefont {Gislason}\ \emph {et~al.}(1985)\citenamefont
  {Gislason}, \citenamefont {Sabelli},\ and\ \citenamefont
  {Wood}}]{PhysRevA.31.2078}%
  \BibitemOpen
  \bibfield  {author} {\bibinfo {author} {\bibfnamefont {Eric~A.}\ \bibnamefont
  {Gislason}}, \bibinfo {author} {\bibfnamefont {Nora~H.}\ \bibnamefont
  {Sabelli}}, \ and\ \bibinfo {author} {\bibfnamefont {John~W.}\ \bibnamefont
  {Wood}},\ }\bibfield  {title} {\enquote {\bibinfo {title} {New form of the
  time-energy uncertainty relation},}\ }\href {\doibase
  10.1103/PhysRevA.31.2078} {\bibfield  {journal} {\bibinfo  {journal} {Phys.
  Rev. A}\ }\textbf {\bibinfo {volume} {31}},\ \bibinfo {pages} {2078--2081}
  (\bibinfo {year} {1985})}\BibitemShut {NoStop}%
\bibitem [{\citenamefont {Boykin}\ \emph {et~al.}(2007)\citenamefont {Boykin},
  \citenamefont {Kharche},\ and\ \citenamefont {Klimeck}}]{2007Timothy}%
  \BibitemOpen
  \bibfield  {author} {\bibinfo {author} {\bibfnamefont {Timothy~B}\
  \bibnamefont {Boykin}}, \bibinfo {author} {\bibfnamefont {Neerav}\
  \bibnamefont {Kharche}}, \ and\ \bibinfo {author} {\bibfnamefont {Gerhard}\
  \bibnamefont {Klimeck}},\ }\bibfield  {title} {\enquote {\bibinfo {title}
  {Evolution time and energy uncertainty},}\ }\href {\doibase
  10.1088/0143-0807/28/4/007} {\bibfield  {journal} {\bibinfo  {journal} {Eur.
  J. Phys.}\ }\textbf {\bibinfo {volume} {28}},\ \bibinfo {pages} {673--678}
  (\bibinfo {year} {2007})}\BibitemShut {NoStop}%
\bibitem [{\citenamefont {Deffner}\ and\ \citenamefont
  {Campbell}(2017)}]{2017Sebastian}%
  \BibitemOpen
  \bibfield  {author} {\bibinfo {author} {\bibfnamefont {Sebastian}\
  \bibnamefont {Deffner}}\ and\ \bibinfo {author} {\bibfnamefont {Steve}\
  \bibnamefont {Campbell}},\ }\bibfield  {title} {\enquote {\bibinfo {title}
  {Quantum speed limits: from heisenberg's uncertainty principle to optimal
  quantum control},}\ }\href {\doibase 10.1088/1751-8121/aa86c6} {\bibfield
  {journal} {\bibinfo  {journal} {J. Phys. A: Math. Theor.}\ }\textbf {\bibinfo
  {volume} {50}},\ \bibinfo {pages} {453001} (\bibinfo {year}
  {2017})}\BibitemShut {NoStop}%
\bibitem [{\citenamefont {Margolus}\ and\ \citenamefont
  {Levitin}(1998)}]{MARGOLUS1998188}%
  \BibitemOpen
  \bibfield  {author} {\bibinfo {author} {\bibfnamefont {Norman}\ \bibnamefont
  {Margolus}}\ and\ \bibinfo {author} {\bibfnamefont {Lev~B.}\ \bibnamefont
  {Levitin}},\ }\bibfield  {title} {\enquote {\bibinfo {title} {The maximum
  speed of dynamical evolution},}\ }\href {\doibase
  https://doi.org/10.1016/S0167-2789(98)00054-2} {\bibfield  {journal}
  {\bibinfo  {journal} {Physica D: Nonlinear Phenomena}\ }\textbf {\bibinfo
  {volume} {120}},\ \bibinfo {pages} {188--195} (\bibinfo {year}
  {1998})}\BibitemShut {NoStop}%
\bibitem [{\citenamefont {Jones}\ and\ \citenamefont
  {Kok}(2010)}]{PhysRevA.82.022107}%
  \BibitemOpen
  \bibfield  {author} {\bibinfo {author} {\bibfnamefont {Philip~J.}\
  \bibnamefont {Jones}}\ and\ \bibinfo {author} {\bibfnamefont {Pieter}\
  \bibnamefont {Kok}},\ }\bibfield  {title} {\enquote {\bibinfo {title}
  {Geometric derivation of the quantum speed limit},}\ }\href {\doibase
  10.1103/PhysRevA.82.022107} {\bibfield  {journal} {\bibinfo  {journal} {Phys.
  Rev. A}\ }\textbf {\bibinfo {volume} {82}},\ \bibinfo {pages} {022107}
  (\bibinfo {year} {2010})}\BibitemShut {NoStop}%
\bibitem [{\citenamefont {Sun}\ and\ \citenamefont
  {Zheng}(2019)}]{PhysRevLett.123.180403}%
  \BibitemOpen
  \bibfield  {author} {\bibinfo {author} {\bibfnamefont {Shuning}\ \bibnamefont
  {Sun}}\ and\ \bibinfo {author} {\bibfnamefont {Yujun}\ \bibnamefont
  {Zheng}},\ }\bibfield  {title} {\enquote {\bibinfo {title} {Distinct bound of
  the quantum speed limit via the gauge invariant distance},}\ }\href {\doibase
  10.1103/PhysRevLett.123.180403} {\bibfield  {journal} {\bibinfo  {journal}
  {Phys. Rev. Lett.}\ }\textbf {\bibinfo {volume} {123}},\ \bibinfo {pages}
  {180403} (\bibinfo {year} {2019})}\BibitemShut {NoStop}%
\bibitem [{\citenamefont {Sun}\ \emph {et~al.}(2021)\citenamefont {Sun},
  \citenamefont {Peng}, \citenamefont {Hu},\ and\ \citenamefont
  {Zheng}}]{PhysRevLett.127.100404}%
  \BibitemOpen
  \bibfield  {author} {\bibinfo {author} {\bibfnamefont {Shuning}\ \bibnamefont
  {Sun}}, \bibinfo {author} {\bibfnamefont {Yonggang}\ \bibnamefont {Peng}},
  \bibinfo {author} {\bibfnamefont {Xianghong}\ \bibnamefont {Hu}}, \ and\
  \bibinfo {author} {\bibfnamefont {Yujun}\ \bibnamefont {Zheng}},\ }\bibfield
  {title} {\enquote {\bibinfo {title} {Quantum speed limit quantified by the
  changing rate of phase},}\ }\href {\doibase 10.1103/PhysRevLett.127.100404}
  {\bibfield  {journal} {\bibinfo  {journal} {Phys. Rev. Lett.}\ }\textbf
  {\bibinfo {volume} {127}},\ \bibinfo {pages} {100404} (\bibinfo {year}
  {2021})}\BibitemShut {NoStop}%
\bibitem [{\citenamefont {Deffner}\ and\ \citenamefont
  {Lutz}(2013)}]{PhysRevLett.111.010402}%
  \BibitemOpen
  \bibfield  {author} {\bibinfo {author} {\bibfnamefont {Sebastian}\
  \bibnamefont {Deffner}}\ and\ \bibinfo {author} {\bibfnamefont {Eric}\
  \bibnamefont {Lutz}},\ }\bibfield  {title} {\enquote {\bibinfo {title}
  {Quantum speed limit for non-markovian dynamics},}\ }\href {\doibase
  10.1103/PhysRevLett.111.010402} {\bibfield  {journal} {\bibinfo  {journal}
  {Phys. Rev. Lett.}\ }\textbf {\bibinfo {volume} {111}},\ \bibinfo {pages}
  {010402} (\bibinfo {year} {2013})}\BibitemShut {NoStop}%
\bibitem [{\citenamefont {Taddei}\ \emph {et~al.}(2013)\citenamefont {Taddei},
  \citenamefont {Escher}, \citenamefont {Davidovich},\ and\ \citenamefont
  {de~Matos~Filho}}]{PhysRevLett.110.050402}%
  \BibitemOpen
  \bibfield  {author} {\bibinfo {author} {\bibfnamefont {M.~M.}\ \bibnamefont
  {Taddei}}, \bibinfo {author} {\bibfnamefont {B.~M.}\ \bibnamefont {Escher}},
  \bibinfo {author} {\bibfnamefont {L.}~\bibnamefont {Davidovich}}, \ and\
  \bibinfo {author} {\bibfnamefont {R.~L.}\ \bibnamefont {de~Matos~Filho}},\
  }\bibfield  {title} {\enquote {\bibinfo {title} {Quantum speed limit for
  physical processes},}\ }\href {\doibase 10.1103/PhysRevLett.110.050402}
  {\bibfield  {journal} {\bibinfo  {journal} {Phys. Rev. Lett.}\ }\textbf
  {\bibinfo {volume} {110}},\ \bibinfo {pages} {050402} (\bibinfo {year}
  {2013})}\BibitemShut {NoStop}%
\bibitem [{\citenamefont {del Campo}\ \emph {et~al.}(2013)\citenamefont {del
  Campo}, \citenamefont {Egusquiza}, \citenamefont {Plenio},\ and\
  \citenamefont {Huelga}}]{PhysRevLett.110.050403}%
  \BibitemOpen
  \bibfield  {author} {\bibinfo {author} {\bibfnamefont {A.}~\bibnamefont {del
  Campo}}, \bibinfo {author} {\bibfnamefont {I.~L.}\ \bibnamefont {Egusquiza}},
  \bibinfo {author} {\bibfnamefont {M.~B.}\ \bibnamefont {Plenio}}, \ and\
  \bibinfo {author} {\bibfnamefont {S.~F.}\ \bibnamefont {Huelga}},\ }\bibfield
   {title} {\enquote {\bibinfo {title} {Quantum speed limits in open system
  dynamics},}\ }\href {\doibase 10.1103/PhysRevLett.110.050403} {\bibfield
  {journal} {\bibinfo  {journal} {Phys. Rev. Lett.}\ }\textbf {\bibinfo
  {volume} {110}},\ \bibinfo {pages} {050403} (\bibinfo {year}
  {2013})}\BibitemShut {NoStop}%
\bibitem [{\citenamefont {Marvian}\ and\ \citenamefont
  {Lidar}(2015)}]{PhysRevLett.115.210402}%
  \BibitemOpen
  \bibfield  {author} {\bibinfo {author} {\bibfnamefont {Iman}\ \bibnamefont
  {Marvian}}\ and\ \bibinfo {author} {\bibfnamefont {Daniel~A.}\ \bibnamefont
  {Lidar}},\ }\bibfield  {title} {\enquote {\bibinfo {title} {Quantum speed
  limits for leakage and decoherence},}\ }\href {\doibase
  10.1103/PhysRevLett.115.210402} {\bibfield  {journal} {\bibinfo  {journal}
  {Phys. Rev. Lett.}\ }\textbf {\bibinfo {volume} {115}},\ \bibinfo {pages}
  {210402} (\bibinfo {year} {2015})}\BibitemShut {NoStop}%
\bibitem [{\citenamefont {Van~Vu}\ and\ \citenamefont
  {Hasegawa}(2021)}]{PhysRevLett.126.010601}%
  \BibitemOpen
  \bibfield  {author} {\bibinfo {author} {\bibfnamefont {Tan}\ \bibnamefont
  {Van~Vu}}\ and\ \bibinfo {author} {\bibfnamefont {Yoshihiko}\ \bibnamefont
  {Hasegawa}},\ }\bibfield  {title} {\enquote {\bibinfo {title} {Geometrical
  bounds of the irreversibility in markovian systems},}\ }\href {\doibase
  10.1103/PhysRevLett.126.010601} {\bibfield  {journal} {\bibinfo  {journal}
  {Phys. Rev. Lett.}\ }\textbf {\bibinfo {volume} {126}},\ \bibinfo {pages}
  {010601} (\bibinfo {year} {2021})}\BibitemShut {NoStop}%
\bibitem [{\citenamefont {Funo}\ \emph {et~al.}(2019)\citenamefont {Funo},
  \citenamefont {Shiraishi},\ and\ \citenamefont {Saito}}]{2019Ken}%
  \BibitemOpen
  \bibfield  {author} {\bibinfo {author} {\bibfnamefont {Ken}\ \bibnamefont
  {Funo}}, \bibinfo {author} {\bibfnamefont {Naoto}\ \bibnamefont {Shiraishi}},
  \ and\ \bibinfo {author} {\bibfnamefont {Keiji}\ \bibnamefont {Saito}},\
  }\bibfield  {title} {\enquote {\bibinfo {title} {Speed limit for open quantum
  systems},}\ }\href {\doibase 10.1088/1367-2630/aaf9f5} {\bibfield  {journal}
  {\bibinfo  {journal} {New J. Phys.}\ }\textbf {\bibinfo {volume} {21}},\
  \bibinfo {pages} {013006} (\bibinfo {year} {2019})}\BibitemShut {NoStop}%
\bibitem [{\citenamefont {Pires}\ \emph {et~al.}(2016)\citenamefont {Pires},
  \citenamefont {Cianciaruso}, \citenamefont {C\'eleri}, \citenamefont
  {Adesso},\ and\ \citenamefont {Soares-Pinto}}]{PhysRevX.6.021031}%
  \BibitemOpen
  \bibfield  {author} {\bibinfo {author} {\bibfnamefont {Diego~Paiva}\
  \bibnamefont {Pires}}, \bibinfo {author} {\bibfnamefont {Marco}\ \bibnamefont
  {Cianciaruso}}, \bibinfo {author} {\bibfnamefont {Lucas~C.}\ \bibnamefont
  {C\'eleri}}, \bibinfo {author} {\bibfnamefont {Gerardo}\ \bibnamefont
  {Adesso}}, \ and\ \bibinfo {author} {\bibfnamefont {Diogo~O.}\ \bibnamefont
  {Soares-Pinto}},\ }\bibfield  {title} {\enquote {\bibinfo {title}
  {Generalized geometric quantum speed limits},}\ }\href {\doibase
  10.1103/PhysRevX.6.021031} {\bibfield  {journal} {\bibinfo  {journal} {Phys.
  Rev. X}\ }\textbf {\bibinfo {volume} {6}},\ \bibinfo {pages} {021031}
  (\bibinfo {year} {2016})}\BibitemShut {NoStop}%
\bibitem [{\citenamefont {Campaioli}\ \emph {et~al.}(2017)\citenamefont
  {Campaioli}, \citenamefont {Pollock}, \citenamefont {Binder}, \citenamefont
  {C\'eleri}, \citenamefont {Goold}, \citenamefont {Vinjanampathy},\ and\
  \citenamefont {Modi}}]{PhysRevLett.118.150601}%
  \BibitemOpen
  \bibfield  {author} {\bibinfo {author} {\bibfnamefont {Francesco}\
  \bibnamefont {Campaioli}}, \bibinfo {author} {\bibfnamefont {Felix~A.}\
  \bibnamefont {Pollock}}, \bibinfo {author} {\bibfnamefont {Felix~C.}\
  \bibnamefont {Binder}}, \bibinfo {author} {\bibfnamefont {Lucas}\
  \bibnamefont {C\'eleri}}, \bibinfo {author} {\bibfnamefont {John}\
  \bibnamefont {Goold}}, \bibinfo {author} {\bibfnamefont {Sai}\ \bibnamefont
  {Vinjanampathy}}, \ and\ \bibinfo {author} {\bibfnamefont {Kavan}\
  \bibnamefont {Modi}},\ }\bibfield  {title} {\enquote {\bibinfo {title}
  {Enhancing the charging power of quantum batteries},}\ }\href {\doibase
  10.1103/PhysRevLett.118.150601} {\bibfield  {journal} {\bibinfo  {journal}
  {Phys. Rev. Lett.}\ }\textbf {\bibinfo {volume} {118}},\ \bibinfo {pages}
  {150601} (\bibinfo {year} {2017})}\BibitemShut {NoStop}%
\bibitem [{\citenamefont {Santos}\ \emph {et~al.}(2019)\citenamefont {Santos},
  \citenamefont {\ifmmode~\mbox{\c{C}}\else \c{C}\fi{}akmak}, \citenamefont
  {Campbell},\ and\ \citenamefont {Zinner}}]{PhysRevE.100.032107}%
  \BibitemOpen
  \bibfield  {author} {\bibinfo {author} {\bibfnamefont {Alan~C.}\ \bibnamefont
  {Santos}}, \bibinfo {author} {\bibfnamefont {Bar\ifmmode \imath \else \i
  \fi{}\ifmmode \mbox{\c{s}}\else~\c{s}\fi{}}\ \bibnamefont
  {\ifmmode~\mbox{\c{C}}\else \c{C}\fi{}akmak}}, \bibinfo {author}
  {\bibfnamefont {Steve}\ \bibnamefont {Campbell}}, \ and\ \bibinfo {author}
  {\bibfnamefont {Nikolaj~T.}\ \bibnamefont {Zinner}},\ }\bibfield  {title}
  {\enquote {\bibinfo {title} {Stable adiabatic quantum batteries},}\ }\href
  {\doibase 10.1103/PhysRevE.100.032107} {\bibfield  {journal} {\bibinfo
  {journal} {Phys. Rev. E}\ }\textbf {\bibinfo {volume} {100}},\ \bibinfo
  {pages} {032107} (\bibinfo {year} {2019})}\BibitemShut {NoStop}%
\bibitem [{\citenamefont {Garc\'{\i}a-Pintos}\ \emph
  {et~al.}(2020)\citenamefont {Garc\'{\i}a-Pintos}, \citenamefont {Hamma},\
  and\ \citenamefont {del Campo}}]{PhysRevLett.125.040601}%
  \BibitemOpen
  \bibfield  {author} {\bibinfo {author} {\bibfnamefont {Luis~Pedro}\
  \bibnamefont {Garc\'{\i}a-Pintos}}, \bibinfo {author} {\bibfnamefont
  {Alioscia}\ \bibnamefont {Hamma}}, \ and\ \bibinfo {author} {\bibfnamefont
  {Adolfo}\ \bibnamefont {del Campo}},\ }\bibfield  {title} {\enquote {\bibinfo
  {title} {Fluctuations in extractable work bound the charging power of quantum
  batteries},}\ }\href {\doibase 10.1103/PhysRevLett.125.040601} {\bibfield
  {journal} {\bibinfo  {journal} {Phys. Rev. Lett.}\ }\textbf {\bibinfo
  {volume} {125}},\ \bibinfo {pages} {040601} (\bibinfo {year}
  {2020})}\BibitemShut {NoStop}%
\bibitem [{\citenamefont {Ashhab}\ \emph {et~al.}(2012)\citenamefont {Ashhab},
  \citenamefont {de~Groot},\ and\ \citenamefont {Nori}}]{PhysRevA.85.052327}%
  \BibitemOpen
  \bibfield  {author} {\bibinfo {author} {\bibfnamefont {S.}~\bibnamefont
  {Ashhab}}, \bibinfo {author} {\bibfnamefont {P.~C.}\ \bibnamefont
  {de~Groot}}, \ and\ \bibinfo {author} {\bibfnamefont {Franco}\ \bibnamefont
  {Nori}},\ }\bibfield  {title} {\enquote {\bibinfo {title} {Speed limits for
  quantum gates in multiqubit systems},}\ }\href {\doibase
  10.1103/PhysRevA.85.052327} {\bibfield  {journal} {\bibinfo  {journal} {Phys.
  Rev. A}\ }\textbf {\bibinfo {volume} {85}},\ \bibinfo {pages} {052327}
  (\bibinfo {year} {2012})}\BibitemShut {NoStop}%
\bibitem [{\citenamefont {Neg\^{\i}rneac}\ \emph {et~al.}(2021)\citenamefont
  {Neg\^{\i}rneac}, \citenamefont {Ali}, \citenamefont {Muthusubramanian},
  \citenamefont {Battistel}, \citenamefont {Sagastizabal}, \citenamefont
  {Moreira}, \citenamefont {Marques}, \citenamefont {Vlothuizen}, \citenamefont
  {Beekman}, \citenamefont {Zachariadis}, \citenamefont {Haider}, \citenamefont
  {Bruno},\ and\ \citenamefont {DiCarlo}}]{PhysRevLett.126.220502}%
  \BibitemOpen
  \bibfield  {author} {\bibinfo {author} {\bibfnamefont {V.}~\bibnamefont
  {Neg\^{\i}rneac}}, \bibinfo {author} {\bibfnamefont {H.}~\bibnamefont {Ali}},
  \bibinfo {author} {\bibfnamefont {N.}~\bibnamefont {Muthusubramanian}},
  \bibinfo {author} {\bibfnamefont {F.}~\bibnamefont {Battistel}}, \bibinfo
  {author} {\bibfnamefont {R.}~\bibnamefont {Sagastizabal}}, \bibinfo {author}
  {\bibfnamefont {M.~S.}\ \bibnamefont {Moreira}}, \bibinfo {author}
  {\bibfnamefont {J.~F.}\ \bibnamefont {Marques}}, \bibinfo {author}
  {\bibfnamefont {W.~J.}\ \bibnamefont {Vlothuizen}}, \bibinfo {author}
  {\bibfnamefont {M.}~\bibnamefont {Beekman}}, \bibinfo {author} {\bibfnamefont
  {C.}~\bibnamefont {Zachariadis}}, \bibinfo {author} {\bibfnamefont
  {N.}~\bibnamefont {Haider}}, \bibinfo {author} {\bibfnamefont
  {A.}~\bibnamefont {Bruno}}, \ and\ \bibinfo {author} {\bibfnamefont
  {L.}~\bibnamefont {DiCarlo}},\ }\bibfield  {title} {\enquote {\bibinfo
  {title} {High-fidelity controlled-$z$ gate with maximal intermediate leakage
  operating at the speed limit in a superconducting quantum processor},}\
  }\href {\doibase 10.1103/PhysRevLett.126.220502} {\bibfield  {journal}
  {\bibinfo  {journal} {Phys. Rev. Lett.}\ }\textbf {\bibinfo {volume} {126}},\
  \bibinfo {pages} {220502} (\bibinfo {year} {2021})}\BibitemShut {NoStop}%
\bibitem [{\citenamefont {Deffner}\ and\ \citenamefont
  {Lutz}(2010)}]{PhysRevLett.105.170402}%
  \BibitemOpen
  \bibfield  {author} {\bibinfo {author} {\bibfnamefont {Sebastian}\
  \bibnamefont {Deffner}}\ and\ \bibinfo {author} {\bibfnamefont {Eric}\
  \bibnamefont {Lutz}},\ }\bibfield  {title} {\enquote {\bibinfo {title}
  {Generalized clausius inequality for nonequilibrium quantum processes},}\
  }\href {\doibase 10.1103/PhysRevLett.105.170402} {\bibfield  {journal}
  {\bibinfo  {journal} {Phys. Rev. Lett.}\ }\textbf {\bibinfo {volume} {105}},\
  \bibinfo {pages} {170402} (\bibinfo {year} {2010})}\BibitemShut {NoStop}%
\bibitem [{\citenamefont {Plastina}\ \emph {et~al.}(2014)\citenamefont
  {Plastina}, \citenamefont {Alecce}, \citenamefont {Apollaro}, \citenamefont
  {Falcone}, \citenamefont {Francica}, \citenamefont {Galve}, \citenamefont
  {Lo~Gullo},\ and\ \citenamefont {Zambrini}}]{PhysRevLett.113.260601}%
  \BibitemOpen
  \bibfield  {author} {\bibinfo {author} {\bibfnamefont {F.}~\bibnamefont
  {Plastina}}, \bibinfo {author} {\bibfnamefont {A.}~\bibnamefont {Alecce}},
  \bibinfo {author} {\bibfnamefont {T.~J.~G.}\ \bibnamefont {Apollaro}},
  \bibinfo {author} {\bibfnamefont {G.}~\bibnamefont {Falcone}}, \bibinfo
  {author} {\bibfnamefont {G.}~\bibnamefont {Francica}}, \bibinfo {author}
  {\bibfnamefont {F.}~\bibnamefont {Galve}}, \bibinfo {author} {\bibfnamefont
  {N.}~\bibnamefont {Lo~Gullo}}, \ and\ \bibinfo {author} {\bibfnamefont
  {R.}~\bibnamefont {Zambrini}},\ }\bibfield  {title} {\enquote {\bibinfo
  {title} {Irreversible work and inner friction in quantum thermodynamic
  processes},}\ }\href {\doibase 10.1103/PhysRevLett.113.260601} {\bibfield
  {journal} {\bibinfo  {journal} {Phys. Rev. Lett.}\ }\textbf {\bibinfo
  {volume} {113}},\ \bibinfo {pages} {260601} (\bibinfo {year}
  {2014})}\BibitemShut {NoStop}%
\bibitem [{\citenamefont {Mancino}\ \emph {et~al.}(2018)\citenamefont
  {Mancino}, \citenamefont {Cavina}, \citenamefont {De~Pasquale}, \citenamefont
  {Sbroscia}, \citenamefont {Booth}, \citenamefont {Roccia}, \citenamefont
  {Gianani}, \citenamefont {Giovannetti},\ and\ \citenamefont
  {Barbieri}}]{PhysRevLett.121.160602}%
  \BibitemOpen
  \bibfield  {author} {\bibinfo {author} {\bibfnamefont {Luca}\ \bibnamefont
  {Mancino}}, \bibinfo {author} {\bibfnamefont {Vasco}\ \bibnamefont {Cavina}},
  \bibinfo {author} {\bibfnamefont {Antonella}\ \bibnamefont {De~Pasquale}},
  \bibinfo {author} {\bibfnamefont {Marco}\ \bibnamefont {Sbroscia}}, \bibinfo
  {author} {\bibfnamefont {Robert~I.}\ \bibnamefont {Booth}}, \bibinfo {author}
  {\bibfnamefont {Emanuele}\ \bibnamefont {Roccia}}, \bibinfo {author}
  {\bibfnamefont {Ilaria}\ \bibnamefont {Gianani}}, \bibinfo {author}
  {\bibfnamefont {Vittorio}\ \bibnamefont {Giovannetti}}, \ and\ \bibinfo
  {author} {\bibfnamefont {Marco}\ \bibnamefont {Barbieri}},\ }\bibfield
  {title} {\enquote {\bibinfo {title} {Geometrical bounds on irreversibility in
  open quantum systems},}\ }\href {\doibase 10.1103/PhysRevLett.121.160602}
  {\bibfield  {journal} {\bibinfo  {journal} {Phys. Rev. Lett.}\ }\textbf
  {\bibinfo {volume} {121}},\ \bibinfo {pages} {160602} (\bibinfo {year}
  {2018})}\BibitemShut {NoStop}%
\bibitem [{\citenamefont {Shiraishi}\ \emph {et~al.}(2018)\citenamefont
  {Shiraishi}, \citenamefont {Funo},\ and\ \citenamefont
  {Saito}}]{PhysRevLett.121.070601}%
  \BibitemOpen
  \bibfield  {author} {\bibinfo {author} {\bibfnamefont {Naoto}\ \bibnamefont
  {Shiraishi}}, \bibinfo {author} {\bibfnamefont {Ken}\ \bibnamefont {Funo}}, \
  and\ \bibinfo {author} {\bibfnamefont {Keiji}\ \bibnamefont {Saito}},\
  }\bibfield  {title} {\enquote {\bibinfo {title} {Speed limit for classical
  stochastic processes},}\ }\href {\doibase 10.1103/PhysRevLett.121.070601}
  {\bibfield  {journal} {\bibinfo  {journal} {Phys. Rev. Lett.}\ }\textbf
  {\bibinfo {volume} {121}},\ \bibinfo {pages} {070601} (\bibinfo {year}
  {2018})}\BibitemShut {NoStop}%
\bibitem [{\citenamefont {Nicholson}\ \emph {et~al.}(2020)\citenamefont
  {Nicholson}, \citenamefont {Garc{\'i}a-Pintos}, \citenamefont {del Campo},\
  and\ \citenamefont {Green}}]{Nicholson2020}%
  \BibitemOpen
  \bibfield  {author} {\bibinfo {author} {\bibfnamefont {Schuyler~B.}\
  \bibnamefont {Nicholson}}, \bibinfo {author} {\bibfnamefont {Luis~Pedro}\
  \bibnamefont {Garc{\'i}a-Pintos}}, \bibinfo {author} {\bibfnamefont {Adolfo}\
  \bibnamefont {del Campo}}, \ and\ \bibinfo {author} {\bibfnamefont
  {Jason~R.}\ \bibnamefont {Green}},\ }\bibfield  {title} {\enquote {\bibinfo
  {title} {Time--information uncertainty relations in thermodynamics},}\ }\href
  {\doibase 10.1038/s41567-020-0981-y} {\bibfield  {journal} {\bibinfo
  {journal} {Nature Physics}\ }\textbf {\bibinfo {volume} {16}},\ \bibinfo
  {pages} {1211--1215} (\bibinfo {year} {2020})}\BibitemShut {NoStop}%
\bibitem [{\citenamefont {Lam}\ \emph {et~al.}(2021)\citenamefont {Lam},
  \citenamefont {Peter}, \citenamefont {Groh}, \citenamefont {Alt},
  \citenamefont {Robens}, \citenamefont {Meschede}, \citenamefont {Negretti},
  \citenamefont {Montangero}, \citenamefont {Calarco},\ and\ \citenamefont
  {Alberti}}]{PhysRevX.11.011035}%
  \BibitemOpen
  \bibfield  {author} {\bibinfo {author} {\bibfnamefont {Manolo~R.}\
  \bibnamefont {Lam}}, \bibinfo {author} {\bibfnamefont {Natalie}\ \bibnamefont
  {Peter}}, \bibinfo {author} {\bibfnamefont {Thorsten}\ \bibnamefont {Groh}},
  \bibinfo {author} {\bibfnamefont {Wolfgang}\ \bibnamefont {Alt}}, \bibinfo
  {author} {\bibfnamefont {Carsten}\ \bibnamefont {Robens}}, \bibinfo {author}
  {\bibfnamefont {Dieter}\ \bibnamefont {Meschede}}, \bibinfo {author}
  {\bibfnamefont {Antonio}\ \bibnamefont {Negretti}}, \bibinfo {author}
  {\bibfnamefont {Simone}\ \bibnamefont {Montangero}}, \bibinfo {author}
  {\bibfnamefont {Tommaso}\ \bibnamefont {Calarco}}, \ and\ \bibinfo {author}
  {\bibfnamefont {Andrea}\ \bibnamefont {Alberti}},\ }\bibfield  {title}
  {\enquote {\bibinfo {title} {Demonstration of quantum brachistochrones
  between distant states of an atom},}\ }\href {\doibase
  10.1103/PhysRevX.11.011035} {\bibfield  {journal} {\bibinfo  {journal} {Phys.
  Rev. X}\ }\textbf {\bibinfo {volume} {11}},\ \bibinfo {pages} {011035}
  (\bibinfo {year} {2021})}\BibitemShut {NoStop}%
\bibitem [{\citenamefont {Giovannetti}\ \emph {et~al.}(2011)\citenamefont
  {Giovannetti}, \citenamefont {Lloyd},\ and\ \citenamefont
  {Maccone}}]{Giovannetti2011}%
  \BibitemOpen
  \bibfield  {author} {\bibinfo {author} {\bibfnamefont {Vittorio}\
  \bibnamefont {Giovannetti}}, \bibinfo {author} {\bibfnamefont {Seth}\
  \bibnamefont {Lloyd}}, \ and\ \bibinfo {author} {\bibfnamefont {Lorenzo}\
  \bibnamefont {Maccone}},\ }\bibfield  {title} {\enquote {\bibinfo {title}
  {Advances in quantum metrology},}\ }\href {\doibase 10.1038/nphoton.2011.35}
  {\bibfield  {journal} {\bibinfo  {journal} {Nature Photonics}\ }\textbf
  {\bibinfo {volume} {5}},\ \bibinfo {pages} {222--229} (\bibinfo {year}
  {2011})}\BibitemShut {NoStop}%
\bibitem [{\citenamefont {Alipour}\ \emph {et~al.}(2014)\citenamefont
  {Alipour}, \citenamefont {Mehboudi},\ and\ \citenamefont
  {Rezakhani}}]{PhysRevLett.112.120405}%
  \BibitemOpen
  \bibfield  {author} {\bibinfo {author} {\bibfnamefont {S.}~\bibnamefont
  {Alipour}}, \bibinfo {author} {\bibfnamefont {M.}~\bibnamefont {Mehboudi}}, \
  and\ \bibinfo {author} {\bibfnamefont {A.~T.}\ \bibnamefont {Rezakhani}},\
  }\bibfield  {title} {\enquote {\bibinfo {title} {Quantum metrology in open
  systems: Dissipative cram\'er-rao bound},}\ }\href {\doibase
  10.1103/PhysRevLett.112.120405} {\bibfield  {journal} {\bibinfo  {journal}
  {Phys. Rev. Lett.}\ }\textbf {\bibinfo {volume} {112}},\ \bibinfo {pages}
  {120405} (\bibinfo {year} {2014})}\BibitemShut {NoStop}%
\bibitem [{\citenamefont {Beau}\ and\ \citenamefont {del
  Campo}(2017)}]{PhysRevLett.119.010403}%
  \BibitemOpen
  \bibfield  {author} {\bibinfo {author} {\bibfnamefont {M.}~\bibnamefont
  {Beau}}\ and\ \bibinfo {author} {\bibfnamefont {A.}~\bibnamefont {del
  Campo}},\ }\bibfield  {title} {\enquote {\bibinfo {title} {Nonlinear quantum
  metrology of many-body open systems},}\ }\href {\doibase
  10.1103/PhysRevLett.119.010403} {\bibfield  {journal} {\bibinfo  {journal}
  {Phys. Rev. Lett.}\ }\textbf {\bibinfo {volume} {119}},\ \bibinfo {pages}
  {010403} (\bibinfo {year} {2017})}\BibitemShut {NoStop}%
\bibitem [{\citenamefont {Falasco}\ and\ \citenamefont
  {Esposito}(2020)}]{PhysRevLett.125.120604}%
  \BibitemOpen
  \bibfield  {author} {\bibinfo {author} {\bibfnamefont {Gianmaria}\
  \bibnamefont {Falasco}}\ and\ \bibinfo {author} {\bibfnamefont
  {Massimiliano}\ \bibnamefont {Esposito}},\ }\bibfield  {title} {\enquote
  {\bibinfo {title} {Dissipation-time uncertainty relation},}\ }\href {\doibase
  10.1103/PhysRevLett.125.120604} {\bibfield  {journal} {\bibinfo  {journal}
  {Phys. Rev. Lett.}\ }\textbf {\bibinfo {volume} {125}},\ \bibinfo {pages}
  {120604} (\bibinfo {year} {2020})}\BibitemShut {NoStop}%
\bibitem [{\citenamefont {Ito}\ and\ \citenamefont
  {Dechant}(2020)}]{PhysRevX.10.021056}%
  \BibitemOpen
  \bibfield  {author} {\bibinfo {author} {\bibfnamefont {Sosuke}\ \bibnamefont
  {Ito}}\ and\ \bibinfo {author} {\bibfnamefont {Andreas}\ \bibnamefont
  {Dechant}},\ }\bibfield  {title} {\enquote {\bibinfo {title} {Stochastic time
  evolution, information geometry, and the cram\'er-rao bound},}\ }\href
  {\doibase 10.1103/PhysRevX.10.021056} {\bibfield  {journal} {\bibinfo
  {journal} {Phys. Rev. X}\ }\textbf {\bibinfo {volume} {10}},\ \bibinfo
  {pages} {021056} (\bibinfo {year} {2020})}\BibitemShut {NoStop}%
\bibitem [{\citenamefont {Shanahan}\ \emph {et~al.}(2018)\citenamefont
  {Shanahan}, \citenamefont {Chenu}, \citenamefont {Margolus},\ and\
  \citenamefont {del Campo}}]{PhysRevLett.120.070401}%
  \BibitemOpen
  \bibfield  {author} {\bibinfo {author} {\bibfnamefont {B.}~\bibnamefont
  {Shanahan}}, \bibinfo {author} {\bibfnamefont {A.}~\bibnamefont {Chenu}},
  \bibinfo {author} {\bibfnamefont {N.}~\bibnamefont {Margolus}}, \ and\
  \bibinfo {author} {\bibfnamefont {A.}~\bibnamefont {del Campo}},\ }\bibfield
  {title} {\enquote {\bibinfo {title} {Quantum speed limits across the
  quantum-to-classical transition},}\ }\href {\doibase
  10.1103/PhysRevLett.120.070401} {\bibfield  {journal} {\bibinfo  {journal}
  {Phys. Rev. Lett.}\ }\textbf {\bibinfo {volume} {120}},\ \bibinfo {pages}
  {070401} (\bibinfo {year} {2018})}\BibitemShut {NoStop}%
\bibitem [{\citenamefont {Deffner}(2017)}]{Deffner2017}%
  \BibitemOpen
  \bibfield  {author} {\bibinfo {author} {\bibfnamefont {Sebastian}\
  \bibnamefont {Deffner}},\ }\bibfield  {title} {\enquote {\bibinfo {title}
  {Geometric quantum speed limits: a case for wigner phase space},}\ }\href
  {\doibase 10.1088/1367-2630/aa83dc} {\bibfield  {journal} {\bibinfo
  {journal} {New Journal of Physics}\ }\textbf {\bibinfo {volume} {19}},\
  \bibinfo {pages} {103018} (\bibinfo {year} {2017})}\BibitemShut {NoStop}%
\bibitem [{\citenamefont {Campaioli}\ \emph {et~al.}(2018)\citenamefont
  {Campaioli}, \citenamefont {Pollock}, \citenamefont {Binder},\ and\
  \citenamefont {Modi}}]{PhysRevLett.120.060409}%
  \BibitemOpen
  \bibfield  {author} {\bibinfo {author} {\bibfnamefont {Francesco}\
  \bibnamefont {Campaioli}}, \bibinfo {author} {\bibfnamefont {Felix~A.}\
  \bibnamefont {Pollock}}, \bibinfo {author} {\bibfnamefont {Felix~C.}\
  \bibnamefont {Binder}}, \ and\ \bibinfo {author} {\bibfnamefont {Kavan}\
  \bibnamefont {Modi}},\ }\bibfield  {title} {\enquote {\bibinfo {title}
  {Tightening quantum speed limits for almost all states},}\ }\href {\doibase
  10.1103/PhysRevLett.120.060409} {\bibfield  {journal} {\bibinfo  {journal}
  {Phys. Rev. Lett.}\ }\textbf {\bibinfo {volume} {120}},\ \bibinfo {pages}
  {060409} (\bibinfo {year} {2018})}\BibitemShut {NoStop}%
\bibitem [{\citenamefont {Campaioli}\ \emph {et~al.}(2019)\citenamefont
  {Campaioli}, \citenamefont {Pollock},\ and\ \citenamefont
  {Modi}}]{Campaioli2019tightrobust}%
  \BibitemOpen
  \bibfield  {author} {\bibinfo {author} {\bibfnamefont {Francesco}\
  \bibnamefont {Campaioli}}, \bibinfo {author} {\bibfnamefont {Felix~A.}\
  \bibnamefont {Pollock}}, \ and\ \bibinfo {author} {\bibfnamefont {Kavan}\
  \bibnamefont {Modi}},\ }\bibfield  {title} {\enquote {\bibinfo {title}
  {Tight, robust, and feasible quantum speed limits for open dynamics},}\
  }\href {\doibase 10.22331/q-2019-08-05-168} {\bibfield  {journal} {\bibinfo
  {journal} {{Quantum}}\ }\textbf {\bibinfo {volume} {3}},\ \bibinfo {pages}
  {168} (\bibinfo {year} {2019})}\BibitemShut {NoStop}%
\bibitem [{\citenamefont {Liu}\ \emph {et~al.}(2016)\citenamefont {Liu},
  \citenamefont {Yang}, \citenamefont {An},\ and\ \citenamefont
  {Xu}}]{PhysRevA.93.020105}%
  \BibitemOpen
  \bibfield  {author} {\bibinfo {author} {\bibfnamefont {Hai-Bin}\ \bibnamefont
  {Liu}}, \bibinfo {author} {\bibfnamefont {W.~L.}\ \bibnamefont {Yang}},
  \bibinfo {author} {\bibfnamefont {Jun-Hong}\ \bibnamefont {An}}, \ and\
  \bibinfo {author} {\bibfnamefont {Zhen-Yu}\ \bibnamefont {Xu}},\ }\bibfield
  {title} {\enquote {\bibinfo {title} {Mechanism for quantum speedup in open
  quantum systems},}\ }\href {\doibase 10.1103/PhysRevA.93.020105} {\bibfield
  {journal} {\bibinfo  {journal} {Phys. Rev. A}\ }\textbf {\bibinfo {volume}
  {93}},\ \bibinfo {pages} {020105} (\bibinfo {year} {2016})}\BibitemShut
  {NoStop}%
\bibitem [{\citenamefont {Aifer}\ and\ \citenamefont
  {Deffner}(2022)}]{Aifer_2022}%
  \BibitemOpen
  \bibfield  {author} {\bibinfo {author} {\bibfnamefont {Maxwell}\ \bibnamefont
  {Aifer}}\ and\ \bibinfo {author} {\bibfnamefont {Sebastian}\ \bibnamefont
  {Deffner}},\ }\bibfield  {title} {\enquote {\bibinfo {title} {From quantum
  speed limits to energy-efficient quantum gates},}\ }\href {\doibase
  10.1088/1367-2630/ac6821} {\bibfield  {journal} {\bibinfo  {journal} {New
  Journal of Physics}\ }\textbf {\bibinfo {volume} {24}},\ \bibinfo {pages}
  {055002} (\bibinfo {year} {2022})}\BibitemShut {NoStop}%
\bibitem [{\citenamefont {Hu}\ \emph {et~al.}(2020)\citenamefont {Hu},
  \citenamefont {Sun},\ and\ \citenamefont {Zheng}}]{PhysRevA.101.042107}%
  \BibitemOpen
  \bibfield  {author} {\bibinfo {author} {\bibfnamefont {Xianghong}\
  \bibnamefont {Hu}}, \bibinfo {author} {\bibfnamefont {Shuning}\ \bibnamefont
  {Sun}}, \ and\ \bibinfo {author} {\bibfnamefont {Yujun}\ \bibnamefont
  {Zheng}},\ }\bibfield  {title} {\enquote {\bibinfo {title} {Quantum speed
  limit via the trajectory ensemble},}\ }\href {\doibase
  10.1103/PhysRevA.101.042107} {\bibfield  {journal} {\bibinfo  {journal}
  {Phys. Rev. A}\ }\textbf {\bibinfo {volume} {101}},\ \bibinfo {pages}
  {042107} (\bibinfo {year} {2020})}\BibitemShut {NoStop}%
\bibitem [{\citenamefont {Marian}\ and\ \citenamefont
  {Marian}(2021)}]{PhysRevA.103.022221}%
  \BibitemOpen
  \bibfield  {author} {\bibinfo {author} {\bibfnamefont {Paulina}\ \bibnamefont
  {Marian}}\ and\ \bibinfo {author} {\bibfnamefont {Tudor~A.}\ \bibnamefont
  {Marian}},\ }\bibfield  {title} {\enquote {\bibinfo {title} {Quantum speed of
  evolution in a markovian bosonic environment},}\ }\href {\doibase
  10.1103/PhysRevA.103.022221} {\bibfield  {journal} {\bibinfo  {journal}
  {Phys. Rev. A}\ }\textbf {\bibinfo {volume} {103}},\ \bibinfo {pages}
  {022221} (\bibinfo {year} {2021})}\BibitemShut {NoStop}%
\bibitem [{\citenamefont {Xu}\ \emph {et~al.}(2019)\citenamefont {Xu},
  \citenamefont {Zhang},\ and\ \citenamefont {Liu}}]{PhysRevA.100.052305}%
  \BibitemOpen
  \bibfield  {author} {\bibinfo {author} {\bibfnamefont {Kai}\ \bibnamefont
  {Xu}}, \bibinfo {author} {\bibfnamefont {Guo-Feng}\ \bibnamefont {Zhang}}, \
  and\ \bibinfo {author} {\bibfnamefont {Wu-Ming}\ \bibnamefont {Liu}},\
  }\bibfield  {title} {\enquote {\bibinfo {title} {Quantum dynamical speedup in
  correlated noisy channels},}\ }\href {\doibase 10.1103/PhysRevA.100.052305}
  {\bibfield  {journal} {\bibinfo  {journal} {Phys. Rev. A}\ }\textbf {\bibinfo
  {volume} {100}},\ \bibinfo {pages} {052305} (\bibinfo {year}
  {2019})}\BibitemShut {NoStop}%
\bibitem [{\citenamefont {Xu}(2016)}]{2016Xu}%
  \BibitemOpen
  \bibfield  {author} {\bibinfo {author} {\bibfnamefont {Zhen-Yu}\ \bibnamefont
  {Xu}},\ }\bibfield  {title} {\enquote {\bibinfo {title} {Detecting quantum
  speedup in closed and open systems},}\ }\href {\doibase
  10.1088/1367-2630/18/7/073005} {\bibfield  {journal} {\bibinfo  {journal}
  {New J. Phys.}\ }\textbf {\bibinfo {volume} {18}},\ \bibinfo {pages} {073005}
  (\bibinfo {year} {2016})}\BibitemShut {NoStop}%
\bibitem [{\citenamefont {Wu}\ and\ \citenamefont {An}()}]{weiwu}%
  \BibitemOpen
  \bibfield  {author} {\bibinfo {author} {\bibfnamefont {Wei}\ \bibnamefont
  {Wu}}\ and\ \bibinfo {author} {\bibfnamefont {Jun-Hong}\ \bibnamefont {An}},\
  }\href@noop {} {\enquote {\bibinfo {title} {Quantum speed limit from
  quantum-state diffusion method},}\ }\Eprint {http://arxiv.org/abs/2206.00321}
  {arXiv:2206.00321} \BibitemShut {NoStop}%
\bibitem [{\citenamefont {Provost}\ and\ \citenamefont
  {Vallee}(1980)}]{Provost1980}%
  \BibitemOpen
  \bibfield  {author} {\bibinfo {author} {\bibfnamefont {J.~P.}\ \bibnamefont
  {Provost}}\ and\ \bibinfo {author} {\bibfnamefont {G.}~\bibnamefont
  {Vallee}},\ }\bibfield  {title} {\enquote {\bibinfo {title} {Riemannian
  structure on manifolds of quantum states},}\ }\href {\doibase
  10.1007/BF02193559} {\bibfield  {journal} {\bibinfo  {journal}
  {Communications in Mathematical Physics}\ }\textbf {\bibinfo {volume} {76}},\
  \bibinfo {pages} {289--301} (\bibinfo {year} {1980})}\BibitemShut {NoStop}%
\bibitem [{\citenamefont {Carlini}\ \emph {et~al.}(2006)\citenamefont
  {Carlini}, \citenamefont {Hosoya}, \citenamefont {Koike},\ and\ \citenamefont
  {Okudaira}}]{PhysRevLett.96.060503}%
  \BibitemOpen
  \bibfield  {author} {\bibinfo {author} {\bibfnamefont {Alberto}\ \bibnamefont
  {Carlini}}, \bibinfo {author} {\bibfnamefont {Akio}\ \bibnamefont {Hosoya}},
  \bibinfo {author} {\bibfnamefont {Tatsuhiko}\ \bibnamefont {Koike}}, \ and\
  \bibinfo {author} {\bibfnamefont {Yosuke}\ \bibnamefont {Okudaira}},\
  }\bibfield  {title} {\enquote {\bibinfo {title} {Time-optimal quantum
  evolution},}\ }\href {\doibase 10.1103/PhysRevLett.96.060503} {\bibfield
  {journal} {\bibinfo  {journal} {Phys. Rev. Lett.}\ }\textbf {\bibinfo
  {volume} {96}},\ \bibinfo {pages} {060503} (\bibinfo {year}
  {2006})}\BibitemShut {NoStop}%
\bibitem [{\citenamefont {Frydryszak}\ and\ \citenamefont
  {Tkachuk}(2008)}]{PhysRevA.77.014103}%
  \BibitemOpen
  \bibfield  {author} {\bibinfo {author} {\bibfnamefont {A.~M.}\ \bibnamefont
  {Frydryszak}}\ and\ \bibinfo {author} {\bibfnamefont {V.~M.}\ \bibnamefont
  {Tkachuk}},\ }\bibfield  {title} {\enquote {\bibinfo {title} {Quantum
  brachistochrone problem for a spin-1 system in a magnetic field},}\ }\href
  {\doibase 10.1103/PhysRevA.77.014103} {\bibfield  {journal} {\bibinfo
  {journal} {Phys. Rev. A}\ }\textbf {\bibinfo {volume} {77}},\ \bibinfo
  {pages} {014103} (\bibinfo {year} {2008})}\BibitemShut {NoStop}%
\bibitem [{\citenamefont {Hegerfeldt}(2013)}]{PhysRevLett.111.260501}%
  \BibitemOpen
  \bibfield  {author} {\bibinfo {author} {\bibfnamefont {Gerhard~C.}\
  \bibnamefont {Hegerfeldt}},\ }\bibfield  {title} {\enquote {\bibinfo {title}
  {Driving at the quantum speed limit: Optimal control of a two-level
  system},}\ }\href {\doibase 10.1103/PhysRevLett.111.260501} {\bibfield
  {journal} {\bibinfo  {journal} {Phys. Rev. Lett.}\ }\textbf {\bibinfo
  {volume} {111}},\ \bibinfo {pages} {260501} (\bibinfo {year}
  {2013})}\BibitemShut {NoStop}%
\bibitem [{\citenamefont {Jozsa}(1994)}]{doi:10.1080/09500349414552171}%
  \BibitemOpen
  \bibfield  {author} {\bibinfo {author} {\bibfnamefont {Richard}\ \bibnamefont
  {Jozsa}},\ }\bibfield  {title} {\enquote {\bibinfo {title} {Fidelity for
  mixed quantum states},}\ }\href {\doibase 10.1080/09500349414552171}
  {\bibfield  {journal} {\bibinfo  {journal} {Journal of Modern Optics}\
  }\textbf {\bibinfo {volume} {41}},\ \bibinfo {pages} {2315--2323} (\bibinfo
  {year} {1994})}\BibitemShut {NoStop}%
\bibitem [{\citenamefont {Braunstein}\ and\ \citenamefont
  {Caves}(1994)}]{PhysRevLett.72.3439}%
  \BibitemOpen
  \bibfield  {author} {\bibinfo {author} {\bibfnamefont {Samuel~L.}\
  \bibnamefont {Braunstein}}\ and\ \bibinfo {author} {\bibfnamefont
  {Carlton~M.}\ \bibnamefont {Caves}},\ }\bibfield  {title} {\enquote {\bibinfo
  {title} {Statistical distance and the geometry of quantum states},}\ }\href
  {\doibase 10.1103/PhysRevLett.72.3439} {\bibfield  {journal} {\bibinfo
  {journal} {Phys. Rev. Lett.}\ }\textbf {\bibinfo {volume} {72}},\ \bibinfo
  {pages} {3439--3443} (\bibinfo {year} {1994})}\BibitemShut {NoStop}%
\bibitem [{\citenamefont {O'Connor}\ \emph {et~al.}(2021)\citenamefont
  {O'Connor}, \citenamefont {Guarnieri},\ and\ \citenamefont
  {Campbell}}]{PhysRevA.103.022210}%
  \BibitemOpen
  \bibfield  {author} {\bibinfo {author} {\bibfnamefont {Eoin}\ \bibnamefont
  {O'Connor}}, \bibinfo {author} {\bibfnamefont {Giacomo}\ \bibnamefont
  {Guarnieri}}, \ and\ \bibinfo {author} {\bibfnamefont {Steve}\ \bibnamefont
  {Campbell}},\ }\bibfield  {title} {\enquote {\bibinfo {title} {Action quantum
  speed limits},}\ }\href {\doibase 10.1103/PhysRevA.103.022210} {\bibfield
  {journal} {\bibinfo  {journal} {Phys. Rev. A}\ }\textbf {\bibinfo {volume}
  {103}},\ \bibinfo {pages} {022210} (\bibinfo {year} {2021})}\BibitemShut
  {NoStop}%
\bibitem [{\citenamefont {Hörnedal}\ \emph {et~al.}(2022)\citenamefont
  {Hörnedal}, \citenamefont {Allan},\ and\ \citenamefont
  {Sönnerborn}}]{H_rnedal_2022}%
  \BibitemOpen
  \bibfield  {author} {\bibinfo {author} {\bibfnamefont {Niklas}\ \bibnamefont
  {Hörnedal}}, \bibinfo {author} {\bibfnamefont {Dan}\ \bibnamefont {Allan}},
  \ and\ \bibinfo {author} {\bibfnamefont {Ole}\ \bibnamefont {Sönnerborn}},\
  }\bibfield  {title} {\enquote {\bibinfo {title} {Extensions of the
  mandelstam{\textendash}tamm quantum speed limit to systems in mixed
  states},}\ }\href {\doibase 10.1088/1367-2630/ac688a} {\bibfield  {journal}
  {\bibinfo  {journal} {New Journal of Physics}\ }\textbf {\bibinfo {volume}
  {24}},\ \bibinfo {pages} {055004} (\bibinfo {year} {2022})}\BibitemShut
  {NoStop}%
\bibitem [{\citenamefont {Braunstein}\ and\ \citenamefont {van
  Loock}(2005)}]{RevModPhys.77.513}%
  \BibitemOpen
  \bibfield  {author} {\bibinfo {author} {\bibfnamefont {Samuel~L.}\
  \bibnamefont {Braunstein}}\ and\ \bibinfo {author} {\bibfnamefont {Peter}\
  \bibnamefont {van Loock}},\ }\bibfield  {title} {\enquote {\bibinfo {title}
  {Quantum information with continuous variables},}\ }\href {\doibase
  10.1103/RevModPhys.77.513} {\bibfield  {journal} {\bibinfo  {journal} {Rev.
  Mod. Phys.}\ }\textbf {\bibinfo {volume} {77}},\ \bibinfo {pages} {513--577}
  (\bibinfo {year} {2005})}\BibitemShut {NoStop}%
\bibitem [{\citenamefont {{\v{S}}afr{\'{a}}nek}\ \emph
  {et~al.}(2015)\citenamefont {{\v{S}}afr{\'{a}}nek}, \citenamefont {Lee},\
  and\ \citenamefont {Fuentes}}]{_afr_nek_2015}%
  \BibitemOpen
  \bibfield  {author} {\bibinfo {author} {\bibfnamefont {Dominik}\ \bibnamefont
  {{\v{S}}afr{\'{a}}nek}}, \bibinfo {author} {\bibfnamefont {Antony~R}\
  \bibnamefont {Lee}}, \ and\ \bibinfo {author} {\bibfnamefont {Ivette}\
  \bibnamefont {Fuentes}},\ }\bibfield  {title} {\enquote {\bibinfo {title}
  {Quantum parameter estimation using multi-mode gaussian states},}\ }\href
  {\doibase 10.1088/1367-2630/17/7/073016} {\bibfield  {journal} {\bibinfo
  {journal} {New Journal of Physics}\ }\textbf {\bibinfo {volume} {17}},\
  \bibinfo {pages} {073016} (\bibinfo {year} {2015})}\BibitemShut {NoStop}%
\bibitem [{\citenamefont {Scutaru}(1998)}]{Scutaru_1998}%
  \BibitemOpen
  \bibfield  {author} {\bibinfo {author} {\bibfnamefont {H}~\bibnamefont
  {Scutaru}},\ }\bibfield  {title} {\enquote {\bibinfo {title} {Fidelity for
  displaced squeezed thermal states and the oscillator semigroup},}\ }\href
  {\doibase 10.1088/0305-4470/31/15/025} {\bibfield  {journal} {\bibinfo
  {journal} {Journal of Physics A: Mathematical and General}\ }\textbf
  {\bibinfo {volume} {31}},\ \bibinfo {pages} {3659--3663} (\bibinfo {year}
  {1998})}\BibitemShut {NoStop}%
\bibitem [{\citenamefont {Marian}\ and\ \citenamefont
  {Marian}(2012)}]{PhysRevA.86.022340}%
  \BibitemOpen
  \bibfield  {author} {\bibinfo {author} {\bibfnamefont {Paulina}\ \bibnamefont
  {Marian}}\ and\ \bibinfo {author} {\bibfnamefont {Tudor~A.}\ \bibnamefont
  {Marian}},\ }\bibfield  {title} {\enquote {\bibinfo {title} {Uhlmann fidelity
  between two-mode gaussian states},}\ }\href {\doibase
  10.1103/PhysRevA.86.022340} {\bibfield  {journal} {\bibinfo  {journal} {Phys.
  Rev. A}\ }\textbf {\bibinfo {volume} {86}},\ \bibinfo {pages} {022340}
  (\bibinfo {year} {2012})}\BibitemShut {NoStop}%
\bibitem [{\citenamefont {{\v{S}}afr{\'{a}}nek}(2018)}]{_afr_nek_2018}%
  \BibitemOpen
  \bibfield  {author} {\bibinfo {author} {\bibfnamefont {Dominik}\ \bibnamefont
  {{\v{S}}afr{\'{a}}nek}},\ }\bibfield  {title} {\enquote {\bibinfo {title}
  {Estimation of gaussian quantum states},}\ }\href {\doibase
  10.1088/1751-8121/aaf068} {\bibfield  {journal} {\bibinfo  {journal} {Journal
  of Physics A: Mathematical and Theoretical}\ }\textbf {\bibinfo {volume}
  {52}},\ \bibinfo {pages} {035304} (\bibinfo {year} {2018})}\BibitemShut
  {NoStop}%
\bibitem [{\citenamefont {An}\ and\ \citenamefont
  {Zhang}(2007)}]{PhysRevA.76.042127}%
  \BibitemOpen
  \bibfield  {author} {\bibinfo {author} {\bibfnamefont {Jun-Hong}\
  \bibnamefont {An}}\ and\ \bibinfo {author} {\bibfnamefont {Wei-Min}\
  \bibnamefont {Zhang}},\ }\bibfield  {title} {\enquote {\bibinfo {title}
  {Non-markovian entanglement dynamics of noisy continuous-variable quantum
  channels},}\ }\href {\doibase 10.1103/PhysRevA.76.042127} {\bibfield
  {journal} {\bibinfo  {journal} {Phys. Rev. A}\ }\textbf {\bibinfo {volume}
  {76}},\ \bibinfo {pages} {042127} (\bibinfo {year} {2007})}\BibitemShut
  {NoStop}%
\bibitem [{\citenamefont {Yang}\ \emph {et~al.}(2014)\citenamefont {Yang},
  \citenamefont {An}, \citenamefont {Luo}, \citenamefont {Li},\ and\
  \citenamefont {Oh}}]{PhysRevE.90.022122}%
  \BibitemOpen
  \bibfield  {author} {\bibinfo {author} {\bibfnamefont {Chun-Jie}\
  \bibnamefont {Yang}}, \bibinfo {author} {\bibfnamefont {Jun-Hong}\
  \bibnamefont {An}}, \bibinfo {author} {\bibfnamefont {Hong-Gang}\
  \bibnamefont {Luo}}, \bibinfo {author} {\bibfnamefont {Yading}\ \bibnamefont
  {Li}}, \ and\ \bibinfo {author} {\bibfnamefont {C.~H.}\ \bibnamefont {Oh}},\
  }\bibfield  {title} {\enquote {\bibinfo {title} {Canonical versus
  noncanonical equilibration dynamics of open quantum systems},}\ }\href
  {\doibase 10.1103/PhysRevE.90.022122} {\bibfield  {journal} {\bibinfo
  {journal} {Phys. Rev. E}\ }\textbf {\bibinfo {volume} {90}},\ \bibinfo
  {pages} {022122} (\bibinfo {year} {2014})}\BibitemShut {NoStop}%
\bibitem [{\citenamefont {Wu}\ \emph {et~al.}(2021)\citenamefont {Wu},
  \citenamefont {Bai},\ and\ \citenamefont {An}}]{PhysRevA.103.L010601}%
  \BibitemOpen
  \bibfield  {author} {\bibinfo {author} {\bibfnamefont {Wei}\ \bibnamefont
  {Wu}}, \bibinfo {author} {\bibfnamefont {Si-Yuan}\ \bibnamefont {Bai}}, \
  and\ \bibinfo {author} {\bibfnamefont {Jun-Hong}\ \bibnamefont {An}},\
  }\bibfield  {title} {\enquote {\bibinfo {title} {Non-markovian sensing of a
  quantum reservoir},}\ }\href {\doibase 10.1103/PhysRevA.103.L010601}
  {\bibfield  {journal} {\bibinfo  {journal} {Phys. Rev. A}\ }\textbf {\bibinfo
  {volume} {103}},\ \bibinfo {pages} {L010601} (\bibinfo {year}
  {2021})}\BibitemShut {NoStop}%
\bibitem [{\citenamefont {Poggi}\ \emph {et~al.}(2021)\citenamefont {Poggi},
  \citenamefont {Campbell},\ and\ \citenamefont
  {Deffner}}]{PRXQuantum.2.040349}%
  \BibitemOpen
  \bibfield  {author} {\bibinfo {author} {\bibfnamefont {Pablo~M.}\
  \bibnamefont {Poggi}}, \bibinfo {author} {\bibfnamefont {Steve}\ \bibnamefont
  {Campbell}}, \ and\ \bibinfo {author} {\bibfnamefont {Sebastian}\
  \bibnamefont {Deffner}},\ }\bibfield  {title} {\enquote {\bibinfo {title}
  {Diverging quantum speed limits: A herald of classicality},}\ }\href
  {\doibase 10.1103/PRXQuantum.2.040349} {\bibfield  {journal} {\bibinfo
  {journal} {PRX Quantum}\ }\textbf {\bibinfo {volume} {2}},\ \bibinfo {pages}
  {040349} (\bibinfo {year} {2021})}\BibitemShut {NoStop}%
\bibitem [{\citenamefont {Chen}\ \emph {et~al.}(2019)\citenamefont {Chen},
  \citenamefont {Jin},\ and\ \citenamefont {Liu}}]{Chen_2019}%
  \BibitemOpen
  \bibfield  {author} {\bibinfo {author} {\bibfnamefont {Chong}\ \bibnamefont
  {Chen}}, \bibinfo {author} {\bibfnamefont {Liang}\ \bibnamefont {Jin}}, \
  and\ \bibinfo {author} {\bibfnamefont {Ren-Bao}\ \bibnamefont {Liu}},\
  }\bibfield  {title} {\enquote {\bibinfo {title} {Sensitivity of parameter
  estimation near the exceptional point of a non-hermitian system},}\ }\href
  {\doibase 10.1088/1367-2630/ab32ab} {\bibfield  {journal} {\bibinfo
  {journal} {New Journal of Physics}\ }\textbf {\bibinfo {volume} {21}},\
  \bibinfo {pages} {083002} (\bibinfo {year} {2019})}\BibitemShut {NoStop}%
\bibitem [{\citenamefont {Liu}\ and\ \citenamefont {Houck}(2017)}]{Liu2017}%
  \BibitemOpen
  \bibfield  {author} {\bibinfo {author} {\bibfnamefont {Yanbing}\ \bibnamefont
  {Liu}}\ and\ \bibinfo {author} {\bibfnamefont {Andrew~A.}\ \bibnamefont
  {Houck}},\ }\bibfield  {title} {\enquote {\bibinfo {title} {Quantum
  electrodynamics near a photonic bandgap},}\ }\href {\doibase
  10.1038/nphys3834} {\bibfield  {journal} {\bibinfo  {journal} {Nature
  Physics}\ }\textbf {\bibinfo {volume} {13}},\ \bibinfo {pages} {48--52}
  (\bibinfo {year} {2017})}\BibitemShut {NoStop}%
\bibitem [{\citenamefont {Krinner}\ \emph {et~al.}(2018)\citenamefont
  {Krinner}, \citenamefont {Stewart}, \citenamefont {Pazmi{\~{n}}o},
  \citenamefont {Kwon},\ and\ \citenamefont {Schneble}}]{Krinner2018}%
  \BibitemOpen
  \bibfield  {author} {\bibinfo {author} {\bibfnamefont {Ludwig}\ \bibnamefont
  {Krinner}}, \bibinfo {author} {\bibfnamefont {Michael}\ \bibnamefont
  {Stewart}}, \bibinfo {author} {\bibfnamefont {Arturo}\ \bibnamefont
  {Pazmi{\~{n}}o}}, \bibinfo {author} {\bibfnamefont {Joonhyuk}\ \bibnamefont
  {Kwon}}, \ and\ \bibinfo {author} {\bibfnamefont {Dominik}\ \bibnamefont
  {Schneble}},\ }\bibfield  {title} {\enquote {\bibinfo {title} {Spontaneous
  emission of matter waves from a tunable open quantum system},}\ }\href
  {\doibase 10.1038/s41586-018-0348-z} {\bibfield  {journal} {\bibinfo
  {journal} {Nature}\ }\textbf {\bibinfo {volume} {559}},\ \bibinfo {pages}
  {589--592} (\bibinfo {year} {2018})}\BibitemShut {NoStop}%
\bibitem [{\citenamefont {Andersen}\ \emph {et~al.}(2011)\citenamefont
  {Andersen}, \citenamefont {Stobbe}, \citenamefont {S{\o}rensen},\ and\
  \citenamefont {Lodahl}}]{Andersen2011}%
  \BibitemOpen
  \bibfield  {author} {\bibinfo {author} {\bibfnamefont {Mads~Lykke}\
  \bibnamefont {Andersen}}, \bibinfo {author} {\bibfnamefont {S{\o}ren}\
  \bibnamefont {Stobbe}}, \bibinfo {author} {\bibfnamefont
  {Anders~S{\o}ndberg}\ \bibnamefont {S{\o}rensen}}, \ and\ \bibinfo {author}
  {\bibfnamefont {Peter}\ \bibnamefont {Lodahl}},\ }\bibfield  {title}
  {\enquote {\bibinfo {title} {Strongly modified plasmon--matter interaction
  with mesoscopic quantum emitters},}\ }\href {\doibase 10.1038/nphys1870}
  {\bibfield  {journal} {\bibinfo  {journal} {Nature Physics}\ }\textbf
  {\bibinfo {volume} {7}},\ \bibinfo {pages} {215--218} (\bibinfo {year}
  {2011})}\BibitemShut {NoStop}%
\bibitem [{\citenamefont {Yang}\ and\ \citenamefont
  {An}(2017)}]{PhysRevB.95.161408}%
  \BibitemOpen
  \bibfield  {author} {\bibinfo {author} {\bibfnamefont {Chun-Jie}\
  \bibnamefont {Yang}}\ and\ \bibinfo {author} {\bibfnamefont {Jun-Hong}\
  \bibnamefont {An}},\ }\bibfield  {title} {\enquote {\bibinfo {title}
  {Suppressed dissipation of a quantum emitter coupled to surface plasmon
  polaritons},}\ }\href {\doibase 10.1103/PhysRevB.95.161408} {\bibfield
  {journal} {\bibinfo  {journal} {Phys. Rev. B}\ }\textbf {\bibinfo {volume}
  {95}},\ \bibinfo {pages} {161408} (\bibinfo {year} {2017})}\BibitemShut
  {NoStop}%
\bibitem [{\citenamefont {Haikka}\ \emph {et~al.}(2011)\citenamefont {Haikka},
  \citenamefont {McEndoo}, \citenamefont {De~Chiara}, \citenamefont {Palma},\
  and\ \citenamefont {Maniscalco}}]{PhysRevA.84.031602}%
  \BibitemOpen
  \bibfield  {author} {\bibinfo {author} {\bibfnamefont {P.}~\bibnamefont
  {Haikka}}, \bibinfo {author} {\bibfnamefont {S.}~\bibnamefont {McEndoo}},
  \bibinfo {author} {\bibfnamefont {G.}~\bibnamefont {De~Chiara}}, \bibinfo
  {author} {\bibfnamefont {G.~M.}\ \bibnamefont {Palma}}, \ and\ \bibinfo
  {author} {\bibfnamefont {S.}~\bibnamefont {Maniscalco}},\ }\bibfield  {title}
  {\enquote {\bibinfo {title} {Quantifying, characterizing, and controlling
  information flow in ultracold atomic gases},}\ }\href {\doibase
  10.1103/PhysRevA.84.031602} {\bibfield  {journal} {\bibinfo  {journal} {Phys.
  Rev. A}\ }\textbf {\bibinfo {volume} {84}},\ \bibinfo {pages} {031602}
  (\bibinfo {year} {2011})}\BibitemShut {NoStop}%
\end{thebibliography}%

\end{document}